\documentclass[%
reprint,
superscriptaddress,
 amsmath,amssymb,
 aps,
 prb,
]{revtex4-2}

\usepackage[colorlinks, linkcolor=blue, anchorcolor=blue, citecolor=blue, urlcolor=blue]{hyperref} 
\usepackage{graphicx}
\usepackage{float}

\begin{document}

\title{Shear-Mode Raman Imaging of Ferroelectric Switching in Multilayer 3$R$-MoS$_2$}

\author{Yulu~Liu}
\affiliation{National Laboratory of Solid State Microstructures and Department of Physics, Nanjing University, Nanjing 210093, China}

\author{Kenji Watanabe}
\affiliation{Research Center for Electronic and Optical Materials, National Institute for Materials Science, 1-1 Namiki, Tsukuba 305-0044, Japan}

\author{Takashi~Taniguchi}
\affiliation{Research Center for Materials Nanoarchitectonics, National Institute for Materials Science,  1-1 Namiki, Tsukuba 305-0044, Japan}

\author{Xiaoxiang~Xi}
\email{xxi@nju.edu.cn}
\affiliation{National Laboratory of Solid State Microstructures and Department of Physics, Nanjing University, Nanjing 210093, China}
\affiliation{Collaborative Innovation Center of Advanced Microstructures, Nanjing University, Nanjing 210093, China}
\affiliation{Jiangsu Physical Science Research Center, Nanjing 210093, China}

\begin{abstract}
We use shear-mode Raman imaging to track ferroelectric switching in multilayer 3$R$-MoS$_2$. Within a single flake, mechanically segmented regions respond independently and follow distinct pathways. Partially polarized end states indicate that domain walls can reside between selected layer pairs, producing partial stacking transformations. The dwell time of intermediate states varies widely, indicating that pinning sites strongly influence the dynamics. Second-harmonic generation measurements further reveal three characteristic sample-boundary and domain-wall orientations, including a prevalent chiral direction near the zigzag-armchair bisector. These results provide a direct, noninvasive view of domain-wall-mediated switching in a prototypical sliding ferroelectric and identify pinning and exfoliation-created boundaries as key factors governing its dynamics.
\end{abstract}

\maketitle

Interfacial ferroelectricity arises in van der Waals layered crystals, extending ferroelectricity beyond conventional systems where polarization originates from ionic displacements~\cite{Wu2022}. Although the monolayer unit can be nonpolar, inversion-breaking stacking enables interlayer charge transfer and produces an out-of-plane polarization, which can be flipped by relative sliding between adjacent layers, often termed sliding ferroelectricity~\cite{Li2017}. This mechanism operates in few-layer crystals, including WTe$_2$~\cite{Fei2018,Xiao2020}, MoTe$_2$~\cite{Jindal2023}, 3$R$-MoS$_2$~\cite{Meng2022,Yang2022,Liang2022,Yang2024,Liang2025,Liang2025PRX,Ouyang2025}, ReS$_2$~\cite{Wan2022}, and rhombohedral BN~\cite{Wang2024}, as well as in moir\'{e} homobilayers or multilayers~\cite{Stern2021,Yasuda2021,Yasuda2024,McGilly2020,Wang2022,Weston2022,Deb2022,Ko2023,Winkle2024,Zheng2020}. These developments enable ultrathin ferroelectric devices~\cite{Li2024,Bian2024} and control over optical and electronic properties~\cite{Yang2022,Liang2022,Yang2024,Liang2025,Liang2025PRX,Ouyang2025,Niu2022,Klein2023,Chen2024}.

Despite substantial advances, the microscopic mechanism of ferroelectric switching remains under active investigation. A central question is how an out-of-plane electric field drives interlayer sliding and thus polarization reversal. Recent theoretical studies highlight the essential role of domain walls (DWs), rather than coherent sliding of entire layers~\cite{He2024,Wang2025,Ke2025,Shi2025}. At a DW, reduced symmetry yields off-diagonal components of the Born effective charge, such that an out-of-plane field generates an in-plane driving force that moves the DW and changes the stacking order and polarization~\cite{He2024,Wang2025,Ke2025,Shi2025}. In principle, DW motion can be superlubric~\cite{Ke2025}, consistent with experimentally observed fast, high-endurance switching~\cite{Yasuda2024,Bian2024,Bai2025}. Systematic studies of switching dynamics are needed to test these theories and to identify practical factors that limit performance.

Rhombohedrally stacked 3$R$-MoS$_2$ (hereafter MoS$_2$) is well suited for such studies because the absence of moir\'{e} effects simplifies domain analysis. This environmentally stable polytype is also appealing for large-scale device integration~\cite{Liu2025b}. Photoluminescence and reflection-contrast spectroscopy have resolved switching pathways in few-layer samples by exploiting coupling between ferroelectricity and excitonic states~\cite{Liang2022,Yang2024,Liang2025,Liang2025PRX,Ouyang2025}. As the layer number increases, stacking possibilities proliferate, yielding multiple ferroelectric states useful for multilevel information storage~\cite{Meng2022}, although their identification becomes more challenging. While microscopic imaging of domains and DW dynamics has been reported for various interfacial ferroelectrics~\cite{Deb2022,Weston2022,Ko2023,Winkle2024}, a systematic device-scale study of domain switching in multilayer MoS$_2$, crucial for understanding device performance, has been lacking.

In this Letter, we establish shear-mode Raman spectroscopy and spatial mapping as a direct structural probe of ferroelectric switching in multilayer MoS$_2$. By tracking the stacking degree of freedom, this method visualizes field-driven domain evolution on the device scale and reveals that nominally single flakes are often partitioned into mechanically segmented regions that switch independently. The diverse multilayer switching pathways imply layer-selective stacking transformations, indicating that domain walls fundamentally constrain the accessible end states and thus the total polarization. The data further distinguish cases in which intermediate states are effectively bypassed from those in which strong pinning stabilizes them, and reveal a prevalent domain-wall orientation beyond existing models.

\begin{figure}[t]
\centering
\includegraphics[width=1.0\linewidth]{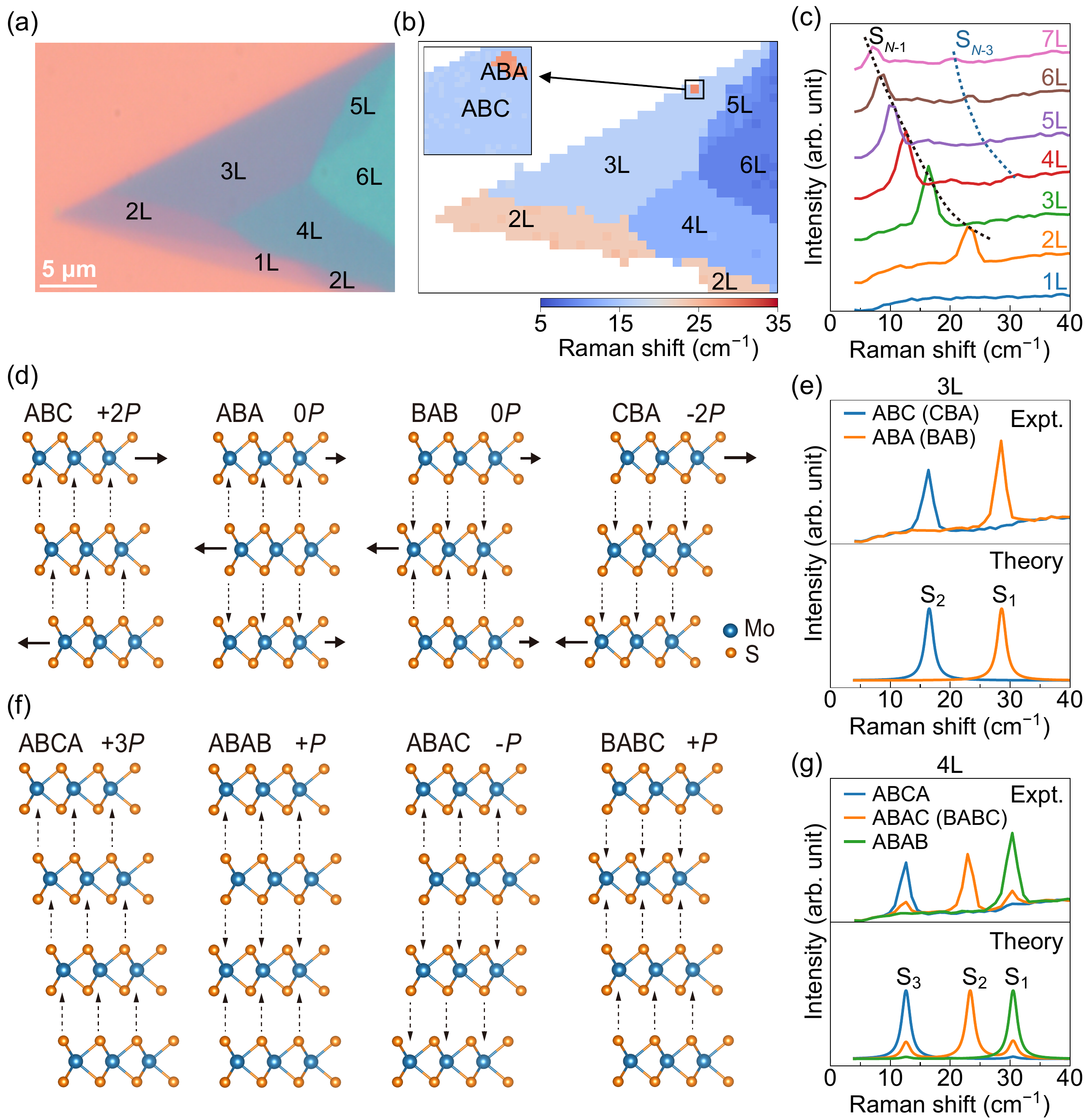}
\caption{(a) Optical micrograph of a MoS$_2$ flake with layer numbers labeled. (b) Corresponding Raman map of the frequency of the most intense shear mode branch. Inset: enlargement of the marked region. (c) Layer-number dependent Raman spectra. S$_{N-1}$ and S$_{N-3}$ denote the two lowest-frequency shear-mode branches. (d),(f) Stacking configurations for trilayer and tetralayer MoS$_2$. Dashed arrows indicate the directions of local electric polarization. Solid arrows in (d) indicate the vibration patterns of shear modes. (e),(g) Measured and calculated stacking-dependent shear mode spectra for trilayer and tetralayer MoS$_2$. }
\label{Fig1}
\end{figure}

\begin{figure*}[t]
\centering
\includegraphics[width=0.8\linewidth]{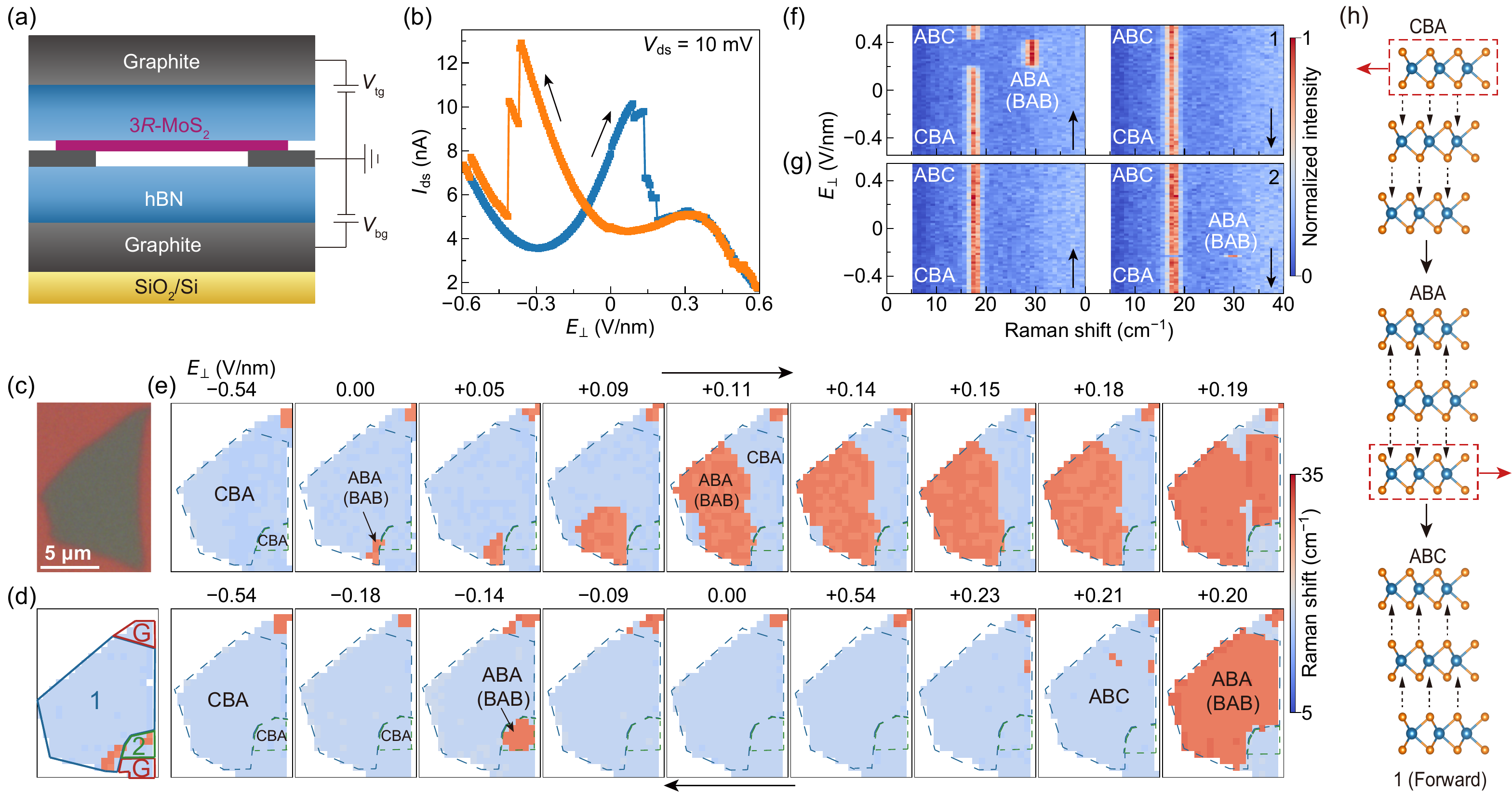}
\caption{(a) Schematic of the dual-gate device. Positive electric field points in the upward direction. (b) Electric-field dependence of the source-drain current in a trilayer device. (c) Optical micrograph of the trilayer flake used in Device D35. (d) The corresponding Raman map of the frequency of the most intense shear mode branch, showing the initial domain distribution, with distinct regions labeled. (e) Raman maps showing the evolution of the domain distribution under a sequence of electric fields. (f),(g) Electric-field-dependent Raman spectra acquired at representative spots in Region~1 (f) and Region~2 (g).  Arrows in (b) and (e)--(g) indicate the field-sweep direction. (h) Illustration of the switching pathway for Region 1 in the forward scan.}
\label{Fig2}
\end{figure*}

The out-of-plane polarization in MoS$_2$ is dictated by the stacking order. Vertical registry of Mo and S atoms in adjacent layers, together with their different electron affinities, produces interlayer charge transfer and thus an electric dipole normal to the layers [Fig.~1(d)]~\cite{Li2017}. An interlayer translation by approximately one Mo-S bond length reverses this registry and flips the polarization~\cite{Liang2022}. In multilayers, the polarizations across successive van der Waals gaps add or cancel~\cite{Deb2022}, so the net polarization depends on the specific stacking configuration, as illustrated in Figs.~1(d) and 1(f).

Raman-active interlayer modes are sensitive to layer number and stacking in van der Waals crystals~\cite{Lu2015,Luo2015,Lee2016,Liang2017,Lin2019,Cong2020,Baren2019}, providing an optical probe of interfacial ferroelectricity. There are two types of interlayer modes, shear modes involving in-plane relative motion [illustrated in Fig.~1(d)] and breathing modes involving out-of-plane motion. Because shear modes reflect the in-plane layer translation associated with sliding ferroelectricity, we selectively measured them in the crossed-polarization configuration~\cite{Baren2019}. Details of the experimental methods can be found in Supplemental Material, Note~1~\footnote{See Supplemental Material for the methods for sample preparation, device fabrication, and optical spectroscopy measurements; the bond polarizability model for stacking-dependent shear modes; additional data for Device D34 and other devices; evidence for a prevalent chiral orientation; pulsed-voltage measurements.}. Within a linear-chain model with identical nearest-neighbor force constants, an $N$-layer rhombohedrally stacked MoS$_2$ hosts shear modes with frequencies $\omega_{j}=\omega_b\cos(j\pi/2N)$ for $j=1,2,\ldots,N-1$, where $\omega_b$ is the bulk-limit frequency~\cite{Baren2019}. In ABC stacking, the $(N-1)$th branch, S$_{N-1}$, is the most intense and was used to map the spatial distribution of layer number in multilayer MoS$_2$ [Fig.~1(a)--1(c)]. The observed redshift of S$_{N-1}$ with increasing $N$ agrees with the model (Supplemental Fig.~2~\cite{Note1}).

The Raman map in Fig.~1(b) is nearly homogeneous within each region of a given thickness, consistent with ABC stacking being the prevalent configuration. Metastable stackings also appear. A small area with a markedly higher frequency for the most intense shear mode branch within an otherwise ABC-stacked trilayer is an ABA or BAB domain. Figure~1(d) illustrates the four trilayer configurations. ABC and CBA are mirror-related about the central layer and carry opposite polarizations. ABA and BAB have zero net polarization and act as intermediate states along the ABC$\leftrightarrow$CBA switching pathway, reached by sliding only the top or the bottom layer. Within the interlayer bond-polarizability model~\cite{Luo2015,Liang2017}, ABC and CBA exhibit only the S$_2$ branch, whereas ABA and BAB exhibit only the S$_1$ branch, with equal Raman intensity within each pair (see Supplemental Material, Note~2~\cite{Note1}). The measured spectra are consistent with these expectations [Fig.~1(e)].

\begin{figure*}[t]
\centering
\includegraphics[width=0.85\linewidth]{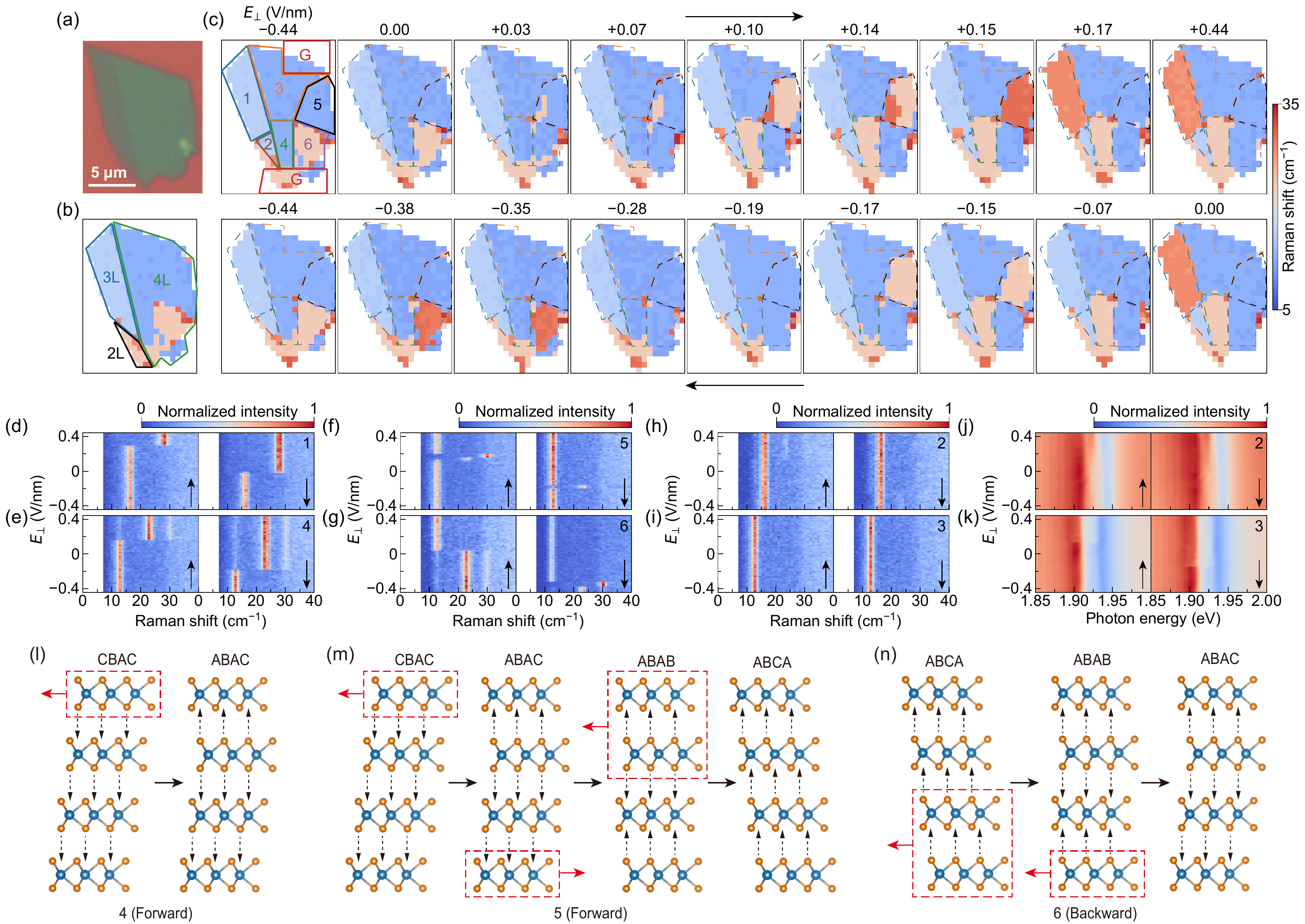}
\caption{(a) Optical micrograph of the multilayer flake used in Device D34. (b) Raman map revealing thickness and initial domain distribution. (c) Raman maps showing the evolution of the domain distribution under a sequence of electric fields. (d)--(i) Electric-field-dependent Raman spectra acquired at representative spots in Regions~1--6, as marked in (c). (j),(k) Corresponding electric-field-dependent reflection-contrast spectra for Regions~2 and 3. Arrows in (c)--(k) indicate the field-sweep direction. (l)--(n) Illustrations of the switching pathway for Regions 4--6 in the forward or backward scan.}
\label{Fig3}
\end{figure*}

This approach extends to thicker samples, as shown for tetralayers in Fig.~1(g) (Supplemental Material, Note~2). Four stacking configurations are depicted in Fig.~1(f), and their mirror-related counterparts are shown in Supplemental Fig.~3. In tetralayers, three shear branches S$_{1\text{--}3}$ appear in order of decreasing frequency with stacking-dependent relative intensities: ABCA and ABAB are dominated by S$_3$ and S$_1$, respectively, with S$_2$ absent; ABAC and BABC share the same response, with an intense S$_2$ and weak S$_1$ and S$_3$. Combined with field-induced switching, configurations with indistinguishable spectra may be further differentiated, constraining the switching pathway. This method complements stacking identification via exciton spectroscopy~\cite{Ouyang2025,Liang2025,Liang2025PRX}. 

To probe ferroelectric switching in MoS$_2$, we fabricated dual-gate field-effect transistors with few-layer flakes as the channel material.  A device schematic is shown in Fig.~2(a). By sweeping the top- and bottom-gate voltages ($V_{\text{tg}}$ and $V_{\text{bg}}$) simultaneously, a vertical electric field $E_\perp= (-V_{\text{tg}}/d_t+ V_{\text{bg}}/d_b)/2$ was applied to the MoS$_2$, where $d_t$ and $d_b$ are the thicknesses of the top and bottom hBN dielectrics, respectively. The MoS$_2$ channel was contacted at its ends by graphite electrodes, allowing measurements of the source-drain current $I_{\mathrm{ds}}$ under a fixed bias  $V_{\mathrm{ds}}$. Figure~2(b) shows $I_{\mathrm{ds}}$ versus $E_\perp$ for a representative trilayer device, exhibiting a butterfly-shaped hysteresis with bistable states near zero field, consistent with ferroelectric switching~\cite{Meng2022}. Abrupt stepwise changes in $I_{\mathrm{ds}}$ at both positive and negative fields indicate sequential flipping of distinct ferroelectric domains.

We tracked the domain-switching process using Raman mapping of the shear-mode frequency. Figure~2(c) shows the optical micrograph of a trilayer flake exfoliated on a SiO$_2$/Si substrate, and Fig.~2(d) presents the corresponding Raman map acquired immediately after the flake was integrated into Device~D35. Guided by the stacking-dependent shear-mode frequencies discussed above, we identify predominantly ABC or CBA stacking across the flake, with small ABA or BAB domains located near the lower-right corner. The presence of multiple domains in as-fabricated devices is essential for ferroelectric switching; devices with a single domain showed no switching within our field window (Supplemental Fig.~4). This is consistent with the theoretical expectation that in the absence of DWs, an out-of-plane field generates no in-plane driving force for interlayer sliding~\cite{Ke2025}.

The response to the applied electric field was region dependent. The two areas overlapping the graphite electrodes [marked G in Fig.~2(d)] did not exhibit ferroelectric switching, presumably due to field screening by the electrodes. The contact-free area was divided into two parts, labeled 1 and 2, which underwent independent switching [Fig.~2(e)]. During the forward scan from $-0.54$ to $0.00$~V/nm, Region~1 retained CBA stacking. A small ABA (or BAB) domain emerged at zero field and expanded as the field was increased. At $0.21$~V/nm, the majority of Region~1 switched abruptly to ABC stacking, leaving only sparse ABA (or BAB) domains that switched at higher fields. This stepwise process is illustrated in Fig.~2(h) as sequential sliding of the top and then the bottom layer. Simultaneous sliding of two layers faces a higher energy barrier~\cite{Meng2022,Fan2025} and is not considered here. The backward scan should follow the reverse sequence with an ABA or BAB intermediate state, depending on which outer layer slides first. However, Region~1 appeared unchanged throughout the backward scan, as also confirmed by measurements at a fixed location [Fig.~2(f)]. Because the switching was reproducible, Region~1 must have recovered CBA at the end of the backward scan, implying that the ABA (or BAB) state was only transient and evaded Raman detection.

At first glance, the switching behavior in Region~2 appears different from that in Region~1, as suggested by the Raman maps [Fig.~2(e)] and field-dependent Raman spectra at a given location [Fig.~2(g)]. However, because the switching in Region~2 was also reproducible, the same CBA$\leftrightarrow$ABA (or BAB)$\leftrightarrow$ABC pathway can be inferred, and the difference from Region~1 lies in the dwell time of intermediate states. In Region~2, the field at which the intermediate state emerged in the backward scan in Figs.~2(e) and 2(g) differs by $\sim 0.10$~V/nm. Such variations in the apparent critical field were frequently observed, indicating that microscopic details of the switching can vary between cycles due to pinning by defects.

To assess the generality of the switching behavior, we performed the same measurements on a sample containing regions with different thickness [Figs.~3(a) and 3(b)]. Because shear-mode spectroscopy does not distinguish AB- and BA-stacked bilayers (Supplemental Fig.~5), we focused on the trilayer and tetralayer portions. Within the tetralayer area, prior to field poling, a triangular domain with ABAC, BABC, or their mirror-symmetric counterparts was identifiable, and ABAB or BABA domains were observed along its perimeter. This sample also showed region-dependent responses to the applied field, with subregions delineated in the first panel of Fig.~3(c). Regions~1--2 lie in the trilayer area, and Regions~3--6 lie in the tetralayer area. Across these regions, we observed four types of seemingly distinct responses, based on Raman maps of the shear-mode frequency together with field-dependent Raman and reflection-contrast spectra at fixed locations [Figs.~3(c)--3(k)].

Regions~1, 4, and 6 exhibit a fully polarized state only at one end of the field sweep. Specifically, Region~1 showed CBA$\rightarrow$ABA (or BAB) switching during the forward scan and recovered CBA during the backward scan. Region~4 likely underwent switching between CBAC and ABAC (or CBAB), involving sliding of only the top (or bottom) layer, as illustrated in Fig.~3(l). In Region~6, several pathways are possible. For example, the backward scan could proceed via ABCA$\rightarrow$ABAB (sliding between the second and third layers), followed by ABAB$\rightarrow$ABAC (sliding of the bottom layer), as illustrated in Fig.~3(n). The forward scan should follow the reverse sequence to recover ABCA stacking, and the apparently missing ABAB state could be due to its short duration, rendering it unresolved by Raman spectroscopy.

Region~5 reached fully polarized states at both ends of the field sweep. One possible pathway is CBAC$\rightarrow$ABAC$\rightarrow$ABAB$\rightarrow$ABCA during the forward scan [Fig.~3(m)], while the backward scan follows the reverse sequence, again with a transient ABAB state evading detection [Fig.~3(f)]. The Raman maps in Fig.~3(c) support this evolution. For completeness, Supplemental Fig.~6 lists alternative switching pathways for Regions~4--6, indistinguishable from the Raman data alone.

The other two response types, observed in the majority of Regions~2 and 3, involved either no discernible switching [Figs.~3(c), 3(h), and 3(j)] or apparent direct ABCA$\leftrightarrow$CBAC switching [Figs.~3(c), 3(i), and 3(k)], neither of which leaves a clear signature in the shear-mode spectra. However, the latter was resolved in reflection-contrast measurements [Fig.~3(k)] as a discrete jump in the spectral range associated with an intralayer exciton. Similar apparent direct ABC$\leftrightarrow$CBA switching was reported in trilayer samples using the same method~\cite{Liang2025}. Exciton spectroscopy also corroborated the switching behavior inferred from shear-mode spectroscopy in Regions~1 and 4--6 (Supplemental Fig.~7). 

Our results show that interlayer sliding driven by DW motion underlies ferroelectric switching. The observed complex switching phenomenology can be distilled into three points, each closely linked to DW properties.

First, for a given thickness the switching pathways are diverse, which we attribute to the number and distribution of active DWs. Bilayer calculations indicate DW widths of order $\sim$10~nm, across which the stacking evolves smoothly from AB to BA~\cite{Ke2025,Shi2025}. Similar stacking DWs (also known as strain solitons or dislocations) have been imaged in multilayer graphene~\cite{Butz2014,Yankowitz2014,Jiang2018,Zhang2022}. Switching between fully polarized end states requires a DW between every adjacent layer pair. Otherwise only a subset of layers can slide, yielding partially polarized or nonpolar end states.  This accounts for the behaviors in Regions~1, 4, and 6 of Device~D34. While surface strain solitons are well documented~\cite{Butz2014,Yankowitz2014,Jiang2018,Zhang2022}, buried solitons mediating internal-layer sliding (as inferred for tetralayer Regions~5 and 6 in Device~D34) merit further study.

\begin{figure}[t]
\centering
\includegraphics[width=\linewidth]{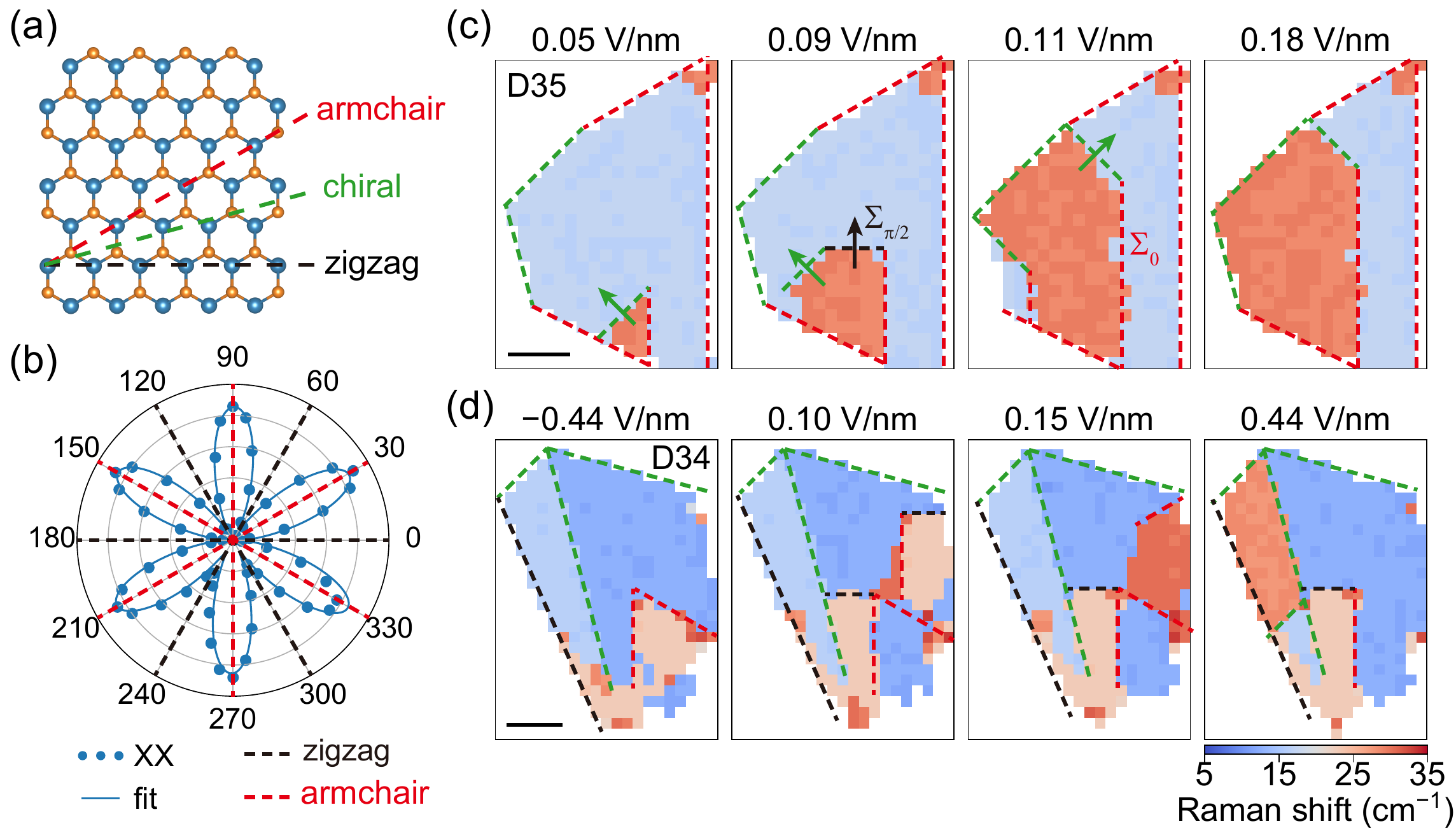}
\caption{(a) Illustration of special directions in the MoS$_2$ lattice. (b) Polarization-angle-dependent SHG intensity for a trilayer in the parallel configuration. The solid line fits $I(\theta)=I_0\sin^2(3\theta)$, where $\theta$ is measured from the zigzag direction. (c),(d) Raman maps in the forward scan for Devices~D35 (c) and D34 (d), selected from Figs.~2 and 3, respectively. Scale bars: 3~$\mu$m. Sample edges, internal-segment boundaries, and DWs are annotated using the line styles defined in (a). Arrows in (c) indicate domain expansion. }
\label{Fig4}
\end{figure}

Second, contiguous flakes split into multiple regions that switch independently and follow distinct pathways (also seen in another trilayer device, Supplemental Fig.~8). Segment boundaries mostly align with the zigzag, armchair, or a direction close to their bisector, as determined by second harmonic generation (SHG, Fig.~4)~\cite{Malard2013}. We attribute these boundaries to strain-induced deformations formed during mechanical exfoliation, which can fracture and evolve into flake edges once a critical strain is exceeded. In Device~D34 (Fig.~3), trilayer and tetralayer portions of the same flake also switched independently. These observations indicate that segment boundaries and thickness steps block DW propagation, likely because the extended DWs cannot conform to such local structural discontinuities. Edges or segment boundaries may also alter DW character, modifying subsequent switching and limiting reproducibility (Supplemental Fig.~9). 

Within the spatial resolution of Raman imaging, three DW orientations are identified in trilayer Device~D35. Following the notation of Ref.~\cite{Ke2025} for rhombohedral BN, two types of DWs along the armchair and zigzag axes are denoted $\Sigma_{0}$ and $\Sigma_{\pi/2}$, for which field-induced sliding is predicted to be parallel and perpendicular to the wall, respectively (see also Ref.~\cite{He2024}). The domain-shape evolution in Fig.~4(c) is consistent with these predictions. A third DW type aligns near the zigzag-armchair bisector, at $(16.3\pm 1.2)^{\circ}$ from the zigzag direction (see analysis in Supplemental Material, Note~4), which lies beyond these models. This orientation can be described by the chiral-angle convention for honeycomb lattices such as in carbon nanotubes, using a chiral vector $\mathbf{C}=n\mathbf{a}_1+m\mathbf{a}_2$, where $(n,m)$ are chiral indices and $\mathbf{a}_1$ and $\mathbf{a}_2$ are basis vectors~\cite{Dresselhaus2001}. Within experimental uncertainty, candidate indices include $(8,3)$, $(7,3)$, and $(5,2)$. This orientation is also prevalent at flake edges and the trilayer-tetralayer step [Figs.~4(c) and 4(d)]. Statistical analysis on a large number of exfoliated flakes further shows that this orientation is prevalent in 3$R$-MoS$_2$ (Supplemental Material, Note~4), in clear contrast to 2$H$-MoS$_2$, where zigzag and armchair edges dominate~\cite{Guo2016}. Its persistence down to the monolayer limit suggests that this preference is associated with exfoliation from the 3$R$ parent crystal, although the microscopic origin remains unresolved.

Third, the field window occupied by intermediate states along the path between fully polarized end states varies widely, as exemplified by Regions~1 and 2 in Device~D35 (Fig.~2) and Regions~3 and 5 in Device~D34 (Fig.~3). The frequent absence of Raman signatures for intermediate states is consistent with superlubric DW motion~\cite{Yeo2025}. Once set in motion, a DW can propagate with velocities up to km$\cdot$s$^{-1}$~\cite{Ke2025,Shi2025,Bai2025,Liang2025PRX}, and may continue advancing even as the field is removed. On the time scale of Raman measurements (seconds per spectrum, i.e. the pixel dwell time in Raman maps), such motion appears instantaneous, rendering intermediate states invisible. Conversely, in regions with pinning the DW velocity can be reduced to $\mu$m$\cdot$s$^{-1}$~\cite{Ko2023} and motion occurs only upon increasing field, stabilizing intermediate states that are then captured by Raman measurements. Pulsed-voltage measurements show that reproducible switching persists down to the instrumental pulse-width limit of 1~$\mu$s, provided that the peak field exceeds the switching threshold (Supplemental Material, Note~5).

In conclusion, we have demonstrated shear-mode Raman spectroscopy as a direct structural probe of ferroelectric switching in multilayer MoS$_2$. The measurements show that polarization reversal is governed by pre-existing domain walls, follows diverse multilayer pathways, and is strongly influenced by mechanical segmentation and pinning. In some regions, domain walls move so rapidly that intermediate states are not resolved, whereas in others, pinning stabilizes intermediate states and sets the critical switching fields. These results establish shear-mode mapping as a powerful approach for directly visualizing domain evolution and for investigating domain-wall physics in interfacial ferroelectrics.

We thank Liangbo Liang, Shi Liu, and Zhicheng Zhong for helpful discussions. This work was supported by the National Key Research and Development Program of China (Grant No.~2024YFA1409100), the Natural Science Foundation of Jiangsu Province (Grants No. BK20231529 and No. BK20233001), the National Natural Science Foundation of China (Grants No. U24A6002 and No. 12474170), and the Fundamental Research Funds for the Central Universities (Grant No. 0204-14380264). K.W. and T.T. acknowledge support from the JSPS KAKENHI (Grant Nos. 20H00354 and 23H02052) and World Premier International Research Center Initiative (WPI), MEXT, Japan.

\textit{Data availability}---The data that support the findings of this article are openly available~\cite{DAS}.


\begin{thebibliography}{53}%
\makeatletter
\providecommand \@ifxundefined [1]{%
 \@ifx{#1\undefined}
}%
\providecommand \@ifnum [1]{%
 \ifnum #1\expandafter \@firstoftwo
 \else \expandafter \@secondoftwo
 \fi
}%
\providecommand \@ifx [1]{%
 \ifx #1\expandafter \@firstoftwo
 \else \expandafter \@secondoftwo
 \fi
}%
\providecommand \natexlab [1]{#1}%
\providecommand \enquote  [1]{``#1''}%
\providecommand \bibnamefont  [1]{#1}%
\providecommand \bibfnamefont [1]{#1}%
\providecommand \citenamefont [1]{#1}%
\providecommand \href@noop [0]{\@secondoftwo}%
\providecommand \href [0]{\begingroup \@sanitize@url \@href}%
\providecommand \@href[1]{\@@startlink{#1}\@@href}%
\providecommand \@@href[1]{\endgroup#1\@@endlink}%
\providecommand \@sanitize@url [0]{\catcode `\\12\catcode `\$12\catcode
  `\&12\catcode `\#12\catcode `\^12\catcode `\_12\catcode `\%12\relax}%
\providecommand \@@startlink[1]{}%
\providecommand \@@endlink[0]{}%
\providecommand \url  [0]{\begingroup\@sanitize@url \@url }%
\providecommand \@url [1]{\endgroup\@href {#1}{\urlprefix }}%
\providecommand \urlprefix  [0]{URL }%
\providecommand \Eprint [0]{\href }%
\providecommand \doibase [0]{https://doi.org/}%
\providecommand \selectlanguage [0]{\@gobble}%
\providecommand \bibinfo  [0]{\@secondoftwo}%
\providecommand \bibfield  [0]{\@secondoftwo}%
\providecommand \translation [1]{[#1]}%
\providecommand \BibitemOpen [0]{}%
\providecommand \bibitemStop [0]{}%
\providecommand \bibitemNoStop [0]{.\EOS\space}%
\providecommand \EOS [0]{\spacefactor3000\relax}%
\providecommand \BibitemShut  [1]{\csname bibitem#1\endcsname}%
\let\auto@bib@innerbib\@empty
\bibitem [{\citenamefont {Wu}\ and\ \citenamefont {Li}(2022)}]{Wu2022}%
  \BibitemOpen
  \bibfield  {author} {\bibinfo {author} {\bibfnamefont {M.~H.}\ \bibnamefont
  {Wu}}\ and\ \bibinfo {author} {\bibfnamefont {J.}~\bibnamefont {Li}},\
  }\bibfield  {title} {\bibinfo {title} {Sliding ferroelectricity in {2D} van
  der {W}aals materials: Related physics and future opportunities},\ }\href
  {https://doi.org/10.1073/pnas.2115703118} {\bibfield  {journal} {\bibinfo
  {journal} {Proc. Natl. Acad. Sci. U.S.A.}\ }\textbf {\bibinfo {volume}
  {118}},\ \bibinfo {pages} {e2115703118} (\bibinfo {year} {2022})}\BibitemShut
  {NoStop}%
\bibitem [{\citenamefont {Li}\ and\ \citenamefont {Wu}(2017)}]{Li2017}%
  \BibitemOpen
  \bibfield  {author} {\bibinfo {author} {\bibfnamefont {L.}~\bibnamefont
  {Li}}\ and\ \bibinfo {author} {\bibfnamefont {M.}~\bibnamefont {Wu}},\
  }\bibfield  {title} {\bibinfo {title} {Binary compound bilayer and multilayer
  with vertical polarizations: Two-dimensional ferroelectrics, multiferroics,
  and nanogenerators},\ }\href {https://doi.org/10.1021/acsnano.7b02756}
  {\bibfield  {journal} {\bibinfo  {journal} {ACS Nano}\ }\textbf {\bibinfo
  {volume} {11}},\ \bibinfo {pages} {6382} (\bibinfo {year}
  {2017})}\BibitemShut {NoStop}%
\bibitem [{\citenamefont {Fei}\ \emph {et~al.}(2018)\citenamefont {Fei},
  \citenamefont {Zhao}, \citenamefont {Palomaki}, \citenamefont {Sun},
  \citenamefont {Miller}, \citenamefont {Zhao}, \citenamefont {Yan},
  \citenamefont {Xu},\ and\ \citenamefont {Cobden}}]{Fei2018}%
  \BibitemOpen
  \bibfield  {author} {\bibinfo {author} {\bibfnamefont {Z.}~\bibnamefont
  {Fei}}, \bibinfo {author} {\bibfnamefont {W.}~\bibnamefont {Zhao}}, \bibinfo
  {author} {\bibfnamefont {T.~A.}\ \bibnamefont {Palomaki}}, \bibinfo {author}
  {\bibfnamefont {B.}~\bibnamefont {Sun}}, \bibinfo {author} {\bibfnamefont
  {M.~K.}\ \bibnamefont {Miller}}, \bibinfo {author} {\bibfnamefont
  {Z.}~\bibnamefont {Zhao}}, \bibinfo {author} {\bibfnamefont {J.}~\bibnamefont
  {Yan}}, \bibinfo {author} {\bibfnamefont {X.}~\bibnamefont {Xu}},\ and\
  \bibinfo {author} {\bibfnamefont {D.~H.}\ \bibnamefont {Cobden}},\ }\bibfield
   {title} {\bibinfo {title} {Ferroelectric switching of a two-dimensional
  metal},\ }\href {https://doi.org/10.1038/s41586-018-0336-3} {\bibfield
  {journal} {\bibinfo  {journal} {Nature}\ }\textbf {\bibinfo {volume} {560}},\
  \bibinfo {pages} {336} (\bibinfo {year} {2018})}\BibitemShut {NoStop}%
\bibitem [{\citenamefont {Xiao}\ \emph {et~al.}(2020)\citenamefont {Xiao},
  \citenamefont {Wang}, \citenamefont {Wang}, \citenamefont {Pemmaraju},
  \citenamefont {Wang}, \citenamefont {Muscher}, \citenamefont {Sie},
  \citenamefont {Nyby}, \citenamefont {Devereaux}, \citenamefont {Qian},
  \citenamefont {Zhang},\ and\ \citenamefont {Lindenberg}}]{Xiao2020}%
  \BibitemOpen
  \bibfield  {author} {\bibinfo {author} {\bibfnamefont {J.}~\bibnamefont
  {Xiao}}, \bibinfo {author} {\bibfnamefont {Y.}~\bibnamefont {Wang}}, \bibinfo
  {author} {\bibfnamefont {H.}~\bibnamefont {Wang}}, \bibinfo {author}
  {\bibfnamefont {C.~D.}\ \bibnamefont {Pemmaraju}}, \bibinfo {author}
  {\bibfnamefont {S.}~\bibnamefont {Wang}}, \bibinfo {author} {\bibfnamefont
  {P.}~\bibnamefont {Muscher}}, \bibinfo {author} {\bibfnamefont {E.~J.}\
  \bibnamefont {Sie}}, \bibinfo {author} {\bibfnamefont {C.~M.}\ \bibnamefont
  {Nyby}}, \bibinfo {author} {\bibfnamefont {T.~P.}\ \bibnamefont {Devereaux}},
  \bibinfo {author} {\bibfnamefont {X.}~\bibnamefont {Qian}}, \bibinfo {author}
  {\bibfnamefont {X.}~\bibnamefont {Zhang}},\ and\ \bibinfo {author}
  {\bibfnamefont {A.~M.}\ \bibnamefont {Lindenberg}},\ }\bibfield  {title}
  {\bibinfo {title} {Berry curvature memory through electrically driven
  stacking transitions},\ }\href {https://doi.org/10.1038/s41567-020-0947-0}
  {\bibfield  {journal} {\bibinfo  {journal} {Nat. Phys.}\ }\textbf {\bibinfo
  {volume} {16}},\ \bibinfo {pages} {1028} (\bibinfo {year}
  {2020})}\BibitemShut {NoStop}%
\bibitem [{\citenamefont {Jindal}\ \emph {et~al.}(2023)\citenamefont {Jindal},
  \citenamefont {Saha}, \citenamefont {Li}, \citenamefont {Taniguchi},
  \citenamefont {Watanabe}, \citenamefont {Hone}, \citenamefont {Birol},
  \citenamefont {Fernandes}, \citenamefont {Dean}, \citenamefont {Pasupathy},\
  and\ \citenamefont {Rhodes}}]{Jindal2023}%
  \BibitemOpen
  \bibfield  {author} {\bibinfo {author} {\bibfnamefont {A.}~\bibnamefont
  {Jindal}}, \bibinfo {author} {\bibfnamefont {A.}~\bibnamefont {Saha}},
  \bibinfo {author} {\bibfnamefont {Z.~Z.}\ \bibnamefont {Li}}, \bibinfo
  {author} {\bibfnamefont {T.}~\bibnamefont {Taniguchi}}, \bibinfo {author}
  {\bibfnamefont {K.}~\bibnamefont {Watanabe}}, \bibinfo {author}
  {\bibfnamefont {J.~C.}\ \bibnamefont {Hone}}, \bibinfo {author}
  {\bibfnamefont {T.}~\bibnamefont {Birol}}, \bibinfo {author} {\bibfnamefont
  {R.~M.}\ \bibnamefont {Fernandes}}, \bibinfo {author} {\bibfnamefont {C.~R.}\
  \bibnamefont {Dean}}, \bibinfo {author} {\bibfnamefont {A.~N.}\ \bibnamefont
  {Pasupathy}},\ and\ \bibinfo {author} {\bibfnamefont {D.~A.}\ \bibnamefont
  {Rhodes}},\ }\bibfield  {title} {\bibinfo {title} {Coupled ferroelectricity
  and superconductivity in bilayer {T$_{\text{d}}$-MoTe$_2$}},\ }\href
  {https://doi.org/10.1038/s41586-022-05521-3} {\bibfield  {journal} {\bibinfo
  {journal} {Nature}\ }\textbf {\bibinfo {volume} {613}},\ \bibinfo {pages}
  {48} (\bibinfo {year} {2023})}\BibitemShut {NoStop}%
\bibitem [{\citenamefont {Meng}\ \emph {et~al.}(2022)\citenamefont {Meng},
  \citenamefont {Wu}, \citenamefont {Bian}, \citenamefont {Pan}, \citenamefont
  {Dong}, \citenamefont {Zhao}, \citenamefont {Chen}, \citenamefont {Wu},
  \citenamefont {Sun}, \citenamefont {Fu}, \citenamefont {Liu}, \citenamefont
  {Shi}, \citenamefont {Zhang}, \citenamefont {Zhang}, \citenamefont {Liu},\
  and\ \citenamefont {Liu}}]{Meng2022}%
  \BibitemOpen
  \bibfield  {author} {\bibinfo {author} {\bibfnamefont {P.}~\bibnamefont
  {Meng}}, \bibinfo {author} {\bibfnamefont {Y.}~\bibnamefont {Wu}}, \bibinfo
  {author} {\bibfnamefont {R.}~\bibnamefont {Bian}}, \bibinfo {author}
  {\bibfnamefont {E.}~\bibnamefont {Pan}}, \bibinfo {author} {\bibfnamefont
  {B.}~\bibnamefont {Dong}}, \bibinfo {author} {\bibfnamefont {X.}~\bibnamefont
  {Zhao}}, \bibinfo {author} {\bibfnamefont {J.}~\bibnamefont {Chen}}, \bibinfo
  {author} {\bibfnamefont {L.}~\bibnamefont {Wu}}, \bibinfo {author}
  {\bibfnamefont {Y.}~\bibnamefont {Sun}}, \bibinfo {author} {\bibfnamefont
  {Q.}~\bibnamefont {Fu}}, \bibinfo {author} {\bibfnamefont {Q.}~\bibnamefont
  {Liu}}, \bibinfo {author} {\bibfnamefont {D.}~\bibnamefont {Shi}}, \bibinfo
  {author} {\bibfnamefont {Q.}~\bibnamefont {Zhang}}, \bibinfo {author}
  {\bibfnamefont {Y.-W.}\ \bibnamefont {Zhang}}, \bibinfo {author}
  {\bibfnamefont {Z.}~\bibnamefont {Liu}},\ and\ \bibinfo {author}
  {\bibfnamefont {F.}~\bibnamefont {Liu}},\ }\bibfield  {title} {\bibinfo
  {title} {Sliding induced multiple polarization states in two-dimensional
  ferroelectrics},\ }\href {https://doi.org/10.1038/s41467-022-35339-6}
  {\bibfield  {journal} {\bibinfo  {journal} {Nat. Commun.}\ }\textbf {\bibinfo
  {volume} {13}},\ \bibinfo {pages} {7696} (\bibinfo {year}
  {2022})}\BibitemShut {NoStop}%
\bibitem [{\citenamefont {Yang}\ \emph {et~al.}(2022)\citenamefont {Yang},
  \citenamefont {Wu}, \citenamefont {Zhou}, \citenamefont {Liang},
  \citenamefont {Ideue}, \citenamefont {Siu}, \citenamefont {Awan},
  \citenamefont {Watanabe}, \citenamefont {Taniguchi}, \citenamefont {Iwasa},
  \citenamefont {Franz},\ and\ \citenamefont {Ye}}]{Yang2022}%
  \BibitemOpen
  \bibfield  {author} {\bibinfo {author} {\bibfnamefont {D.}~\bibnamefont
  {Yang}}, \bibinfo {author} {\bibfnamefont {J.}~\bibnamefont {Wu}}, \bibinfo
  {author} {\bibfnamefont {B.~T.}\ \bibnamefont {Zhou}}, \bibinfo {author}
  {\bibfnamefont {J.}~\bibnamefont {Liang}}, \bibinfo {author} {\bibfnamefont
  {T.}~\bibnamefont {Ideue}}, \bibinfo {author} {\bibfnamefont
  {T.}~\bibnamefont {Siu}}, \bibinfo {author} {\bibfnamefont {K.~M.}\
  \bibnamefont {Awan}}, \bibinfo {author} {\bibfnamefont {K.}~\bibnamefont
  {Watanabe}}, \bibinfo {author} {\bibfnamefont {T.}~\bibnamefont {Taniguchi}},
  \bibinfo {author} {\bibfnamefont {Y.}~\bibnamefont {Iwasa}}, \bibinfo
  {author} {\bibfnamefont {M.}~\bibnamefont {Franz}},\ and\ \bibinfo {author}
  {\bibfnamefont {Z.}~\bibnamefont {Ye}},\ }\bibfield  {title} {\bibinfo
  {title} {Spontaneous-polarization-induced photovoltaic effect in
  rhombohedrally stacked {MoS$_2$}},\ }\href
  {https://doi.org/10.1038/s41566-022-01008-9} {\bibfield  {journal} {\bibinfo
  {journal} {Nat. Photon.}\ }\textbf {\bibinfo {volume} {16}},\ \bibinfo
  {pages} {469} (\bibinfo {year} {2022})}\BibitemShut {NoStop}%
\bibitem [{\citenamefont {Liang}\ \emph {et~al.}(2022)\citenamefont {Liang},
  \citenamefont {Yang}, \citenamefont {Wu}, \citenamefont {Dadap},
  \citenamefont {Watanabe}, \citenamefont {Taniguchi},\ and\ \citenamefont
  {Ye}}]{Liang2022}%
  \BibitemOpen
  \bibfield  {author} {\bibinfo {author} {\bibfnamefont {J.}~\bibnamefont
  {Liang}}, \bibinfo {author} {\bibfnamefont {D.}~\bibnamefont {Yang}},
  \bibinfo {author} {\bibfnamefont {J.}~\bibnamefont {Wu}}, \bibinfo {author}
  {\bibfnamefont {J.}~\bibnamefont {Dadap}}, \bibinfo {author} {\bibfnamefont
  {K.}~\bibnamefont {Watanabe}}, \bibinfo {author} {\bibfnamefont
  {T.}~\bibnamefont {Taniguchi}},\ and\ \bibinfo {author} {\bibfnamefont
  {Z.}~\bibnamefont {Ye}},\ }\bibfield  {title} {\bibinfo {title} {Optically
  probing the asymmetric interlayer coupling in rhombohedral-stacked
  {MoS}$_{\textrm{2}}$ bilayer},\ }\href
  {https://doi.org/10.1103/PhysRevX.12.041005} {\bibfield  {journal} {\bibinfo
  {journal} {Phys. Rev. X}\ }\textbf {\bibinfo {volume} {12}},\ \bibinfo
  {pages} {041005} (\bibinfo {year} {2022})}\BibitemShut {NoStop}%
\bibitem [{\citenamefont {Yang}\ \emph {et~al.}(2024)\citenamefont {Yang},
  \citenamefont {Liang}, \citenamefont {Wu}, \citenamefont {Xiao},
  \citenamefont {Dadap}, \citenamefont {Watanabe}, \citenamefont {Taniguchi},\
  and\ \citenamefont {Ye}}]{Yang2024}%
  \BibitemOpen
  \bibfield  {author} {\bibinfo {author} {\bibfnamefont {D.}~\bibnamefont
  {Yang}}, \bibinfo {author} {\bibfnamefont {J.}~\bibnamefont {Liang}},
  \bibinfo {author} {\bibfnamefont {J.}~\bibnamefont {Wu}}, \bibinfo {author}
  {\bibfnamefont {Y.}~\bibnamefont {Xiao}}, \bibinfo {author} {\bibfnamefont
  {J.~I.}\ \bibnamefont {Dadap}}, \bibinfo {author} {\bibfnamefont
  {K.}~\bibnamefont {Watanabe}}, \bibinfo {author} {\bibfnamefont
  {T.}~\bibnamefont {Taniguchi}},\ and\ \bibinfo {author} {\bibfnamefont
  {Z.}~\bibnamefont {Ye}},\ }\bibfield  {title} {\bibinfo {title} {Non-volatile
  electrical polarization switching via domain wall release in
  {3R}-{MoS}$_{\textrm{2}}$ bilayer},\ }\href
  {https://doi.org/10.1038/s41467-024-45709-x} {\bibfield  {journal} {\bibinfo
  {journal} {Nat. Commun.}\ }\textbf {\bibinfo {volume} {15}},\ \bibinfo
  {pages} {1389} (\bibinfo {year} {2024})}\BibitemShut {NoStop}%
\bibitem [{\citenamefont {Liang}\ \emph
  {et~al.}(2025{\natexlab{a}})\citenamefont {Liang}, \citenamefont {Yang},
  \citenamefont {Wu}, \citenamefont {Xiao}, \citenamefont {Watanabe},
  \citenamefont {Taniguchi}, \citenamefont {Dadap},\ and\ \citenamefont
  {Ye}}]{Liang2025}%
  \BibitemOpen
  \bibfield  {author} {\bibinfo {author} {\bibfnamefont {J.}~\bibnamefont
  {Liang}}, \bibinfo {author} {\bibfnamefont {D.}~\bibnamefont {Yang}},
  \bibinfo {author} {\bibfnamefont {J.}~\bibnamefont {Wu}}, \bibinfo {author}
  {\bibfnamefont {Y.}~\bibnamefont {Xiao}}, \bibinfo {author} {\bibfnamefont
  {K.}~\bibnamefont {Watanabe}}, \bibinfo {author} {\bibfnamefont
  {T.}~\bibnamefont {Taniguchi}}, \bibinfo {author} {\bibfnamefont {J.~I.}\
  \bibnamefont {Dadap}},\ and\ \bibinfo {author} {\bibfnamefont
  {Z.}~\bibnamefont {Ye}},\ }\bibfield  {title} {\bibinfo {title} {Resolving
  polarization switching pathways of sliding ferroelectricity in trilayer
  {3R}-{MoS}$_{\textrm{2}}$},\ }\href
  {https://doi.org/10.1038/s41565-025-01862-y} {\bibfield  {journal} {\bibinfo
  {journal} {Nat. Nanotechnol.}\ }\textbf {\bibinfo {volume} {20}},\ \bibinfo
  {pages} {500} (\bibinfo {year} {2025}{\natexlab{a}})}\BibitemShut {NoStop}%
\bibitem [{\citenamefont {Liang}\ \emph
  {et~al.}(2025{\natexlab{b}})\citenamefont {Liang}, \citenamefont {Xie},
  \citenamefont {Yang}, \citenamefont {Guo}, \citenamefont {Watanabe},
  \citenamefont {Taniguchi}, \citenamefont {Dadap}, \citenamefont {Jones},\
  and\ \citenamefont {Ye}}]{Liang2025PRX}%
  \BibitemOpen
  \bibfield  {author} {\bibinfo {author} {\bibfnamefont {J.}~\bibnamefont
  {Liang}}, \bibinfo {author} {\bibfnamefont {Y.}~\bibnamefont {Xie}}, \bibinfo
  {author} {\bibfnamefont {D.}~\bibnamefont {Yang}}, \bibinfo {author}
  {\bibfnamefont {S.}~\bibnamefont {Guo}}, \bibinfo {author} {\bibfnamefont
  {K.}~\bibnamefont {Watanabe}}, \bibinfo {author} {\bibfnamefont
  {T.}~\bibnamefont {Taniguchi}}, \bibinfo {author} {\bibfnamefont {J.~I.}\
  \bibnamefont {Dadap}}, \bibinfo {author} {\bibfnamefont {D.}~\bibnamefont
  {Jones}},\ and\ \bibinfo {author} {\bibfnamefont {Z.}~\bibnamefont {Ye}},\
  }\bibfield  {title} {\bibinfo {title} {Nanosecond ferroelectric switching of
  intralayer excitons in bilayer {3R}-{MoS}$_{\textrm{2}}$ through coulomb
  engineering},\ }\href {https://doi.org/10.1103/PhysRevX.15.021081} {\bibfield
   {journal} {\bibinfo  {journal} {Phys. Rev. X}\ }\textbf {\bibinfo {volume}
  {15}},\ \bibinfo {pages} {021081} (\bibinfo {year}
  {2025}{\natexlab{b}})}\BibitemShut {NoStop}%
\bibitem [{\citenamefont {Ouyang}\ \emph {et~al.}(2025)\citenamefont {Ouyang},
  \citenamefont {Cha}, \citenamefont {Sun}, \citenamefont {Taniguchi},
  \citenamefont {Watanabe}, \citenamefont {Gabor},\ and\ \citenamefont
  {Lui}}]{Ouyang2025}%
  \BibitemOpen
  \bibfield  {author} {\bibinfo {author} {\bibfnamefont {T.}~\bibnamefont
  {Ouyang}}, \bibinfo {author} {\bibfnamefont {S.}~\bibnamefont {Cha}},
  \bibinfo {author} {\bibfnamefont {Y.}~\bibnamefont {Sun}}, \bibinfo {author}
  {\bibfnamefont {T.}~\bibnamefont {Taniguchi}}, \bibinfo {author}
  {\bibfnamefont {K.}~\bibnamefont {Watanabe}}, \bibinfo {author}
  {\bibfnamefont {N.~M.}\ \bibnamefont {Gabor}},\ and\ \bibinfo {author}
  {\bibfnamefont {C.~H.}\ \bibnamefont {Lui}},\ }\bibfield  {title} {\bibinfo
  {title} {Electrically switching ferroelectric order in
  {3R}-{MoS}$_{\textrm{2}}$ layers},\ }\href
  {https://doi.org/10.1021/acs.nanolett.4c05370} {\bibfield  {journal}
  {\bibinfo  {journal} {Nano Lett.}\ }\textbf {\bibinfo {volume} {25}},\
  \bibinfo {pages} {1459} (\bibinfo {year} {2025})}\BibitemShut {NoStop}%
\bibitem [{\citenamefont {Wan}\ \emph {et~al.}(2022)\citenamefont {Wan},
  \citenamefont {Hu}, \citenamefont {Mao}, \citenamefont {Fu}, \citenamefont
  {Yuan}, \citenamefont {Song}, \citenamefont {Gan}, \citenamefont {Xu},
  \citenamefont {Xue}, \citenamefont {Cheng}, \citenamefont {Huang},
  \citenamefont {Yang}, \citenamefont {Dai}, \citenamefont {Zeng},\ and\
  \citenamefont {Kan}}]{Wan2022}%
  \BibitemOpen
  \bibfield  {author} {\bibinfo {author} {\bibfnamefont {Y.}~\bibnamefont
  {Wan}}, \bibinfo {author} {\bibfnamefont {T.}~\bibnamefont {Hu}}, \bibinfo
  {author} {\bibfnamefont {X.}~\bibnamefont {Mao}}, \bibinfo {author}
  {\bibfnamefont {J.}~\bibnamefont {Fu}}, \bibinfo {author} {\bibfnamefont
  {K.}~\bibnamefont {Yuan}}, \bibinfo {author} {\bibfnamefont {Y.}~\bibnamefont
  {Song}}, \bibinfo {author} {\bibfnamefont {X.}~\bibnamefont {Gan}}, \bibinfo
  {author} {\bibfnamefont {X.}~\bibnamefont {Xu}}, \bibinfo {author}
  {\bibfnamefont {M.}~\bibnamefont {Xue}}, \bibinfo {author} {\bibfnamefont
  {X.}~\bibnamefont {Cheng}}, \bibinfo {author} {\bibfnamefont
  {C.}~\bibnamefont {Huang}}, \bibinfo {author} {\bibfnamefont
  {J.}~\bibnamefont {Yang}}, \bibinfo {author} {\bibfnamefont {L.}~\bibnamefont
  {Dai}}, \bibinfo {author} {\bibfnamefont {H.}~\bibnamefont {Zeng}},\ and\
  \bibinfo {author} {\bibfnamefont {E.}~\bibnamefont {Kan}},\ }\bibfield
  {title} {\bibinfo {title} {Room-temperature ferroelectricity in
  1{T}$^{\prime}$-{ReS}$_2$ multilayers},\ }\href
  {https://doi.org/10.1103/PhysRevLett.128.067601} {\bibfield  {journal}
  {\bibinfo  {journal} {Phys. Rev. Lett.}\ }\textbf {\bibinfo {volume} {128}},\
  \bibinfo {pages} {067601} (\bibinfo {year} {2022})}\BibitemShut {NoStop}%
\bibitem [{\citenamefont {Wang}\ \emph {et~al.}(2024)\citenamefont {Wang},
  \citenamefont {Qi}, \citenamefont {Wei}, \citenamefont {Wu}, \citenamefont
  {Zhang}, \citenamefont {Li}, \citenamefont {Sun}, \citenamefont {Guo},
  \citenamefont {Cao}, \citenamefont {Wang}, \citenamefont {Zhao},
  \citenamefont {Sheng}, \citenamefont {Liu}, \citenamefont {Liu},
  \citenamefont {Wu}, \citenamefont {Xu}, \citenamefont {Wang}, \citenamefont
  {Hong}, \citenamefont {Gao}, \citenamefont {Wu}, \citenamefont {Wang},
  \citenamefont {Xu}, \citenamefont {Wang}, \citenamefont {Ding}, \citenamefont
  {Zheng}, \citenamefont {Liu},\ and\ \citenamefont {Bai}}]{Wang2024}%
  \BibitemOpen
  \bibfield  {author} {\bibinfo {author} {\bibfnamefont {L.}~\bibnamefont
  {Wang}}, \bibinfo {author} {\bibfnamefont {J.}~\bibnamefont {Qi}}, \bibinfo
  {author} {\bibfnamefont {W.}~\bibnamefont {Wei}}, \bibinfo {author}
  {\bibfnamefont {M.}~\bibnamefont {Wu}}, \bibinfo {author} {\bibfnamefont
  {Z.}~\bibnamefont {Zhang}}, \bibinfo {author} {\bibfnamefont
  {X.}~\bibnamefont {Li}}, \bibinfo {author} {\bibfnamefont {H.}~\bibnamefont
  {Sun}}, \bibinfo {author} {\bibfnamefont {Q.}~\bibnamefont {Guo}}, \bibinfo
  {author} {\bibfnamefont {M.}~\bibnamefont {Cao}}, \bibinfo {author}
  {\bibfnamefont {Q.}~\bibnamefont {Wang}}, \bibinfo {author} {\bibfnamefont
  {C.}~\bibnamefont {Zhao}}, \bibinfo {author} {\bibfnamefont {Y.}~\bibnamefont
  {Sheng}}, \bibinfo {author} {\bibfnamefont {Z.}~\bibnamefont {Liu}}, \bibinfo
  {author} {\bibfnamefont {C.}~\bibnamefont {Liu}}, \bibinfo {author}
  {\bibfnamefont {M.}~\bibnamefont {Wu}}, \bibinfo {author} {\bibfnamefont
  {Z.}~\bibnamefont {Xu}}, \bibinfo {author} {\bibfnamefont {W.}~\bibnamefont
  {Wang}}, \bibinfo {author} {\bibfnamefont {H.}~\bibnamefont {Hong}}, \bibinfo
  {author} {\bibfnamefont {P.}~\bibnamefont {Gao}}, \bibinfo {author}
  {\bibfnamefont {M.}~\bibnamefont {Wu}}, \bibinfo {author} {\bibfnamefont
  {Z.-J.}\ \bibnamefont {Wang}}, \bibinfo {author} {\bibfnamefont
  {X.}~\bibnamefont {Xu}}, \bibinfo {author} {\bibfnamefont {E.}~\bibnamefont
  {Wang}}, \bibinfo {author} {\bibfnamefont {F.}~\bibnamefont {Ding}}, \bibinfo
  {author} {\bibfnamefont {X.}~\bibnamefont {Zheng}}, \bibinfo {author}
  {\bibfnamefont {K.}~\bibnamefont {Liu}},\ and\ \bibinfo {author}
  {\bibfnamefont {X.}~\bibnamefont {Bai}},\ }\bibfield  {title} {\bibinfo
  {title} {Bevel-edge epitaxy of ferroelectric rhombohedral boron nitride
  single crystal},\ }\href {https://doi.org/10.1038/s41586-024-07286-3}
  {\bibfield  {journal} {\bibinfo  {journal} {Nature}\ }\textbf {\bibinfo
  {volume} {629}},\ \bibinfo {pages} {74} (\bibinfo {year} {2024})}\BibitemShut
  {NoStop}%
\bibitem [{\citenamefont {Stern}\ \emph {et~al.}(2021)\citenamefont {Stern},
  \citenamefont {Waschitz}, \citenamefont {Cao}, \citenamefont {Nevo},
  \citenamefont {Watanabe}, \citenamefont {Taniguchi}, \citenamefont {Sela},
  \citenamefont {Urbakh}, \citenamefont {Hod},\ and\ \citenamefont
  {Shalom}}]{Stern2021}%
  \BibitemOpen
  \bibfield  {author} {\bibinfo {author} {\bibfnamefont {M.~V.}\ \bibnamefont
  {Stern}}, \bibinfo {author} {\bibfnamefont {Y.}~\bibnamefont {Waschitz}},
  \bibinfo {author} {\bibfnamefont {W.}~\bibnamefont {Cao}}, \bibinfo {author}
  {\bibfnamefont {I.}~\bibnamefont {Nevo}}, \bibinfo {author} {\bibfnamefont
  {K.}~\bibnamefont {Watanabe}}, \bibinfo {author} {\bibfnamefont
  {T.}~\bibnamefont {Taniguchi}}, \bibinfo {author} {\bibfnamefont
  {E.}~\bibnamefont {Sela}}, \bibinfo {author} {\bibfnamefont {M.}~\bibnamefont
  {Urbakh}}, \bibinfo {author} {\bibfnamefont {O.}~\bibnamefont {Hod}},\ and\
  \bibinfo {author} {\bibfnamefont {M.~B.}\ \bibnamefont {Shalom}},\ }\bibfield
   {title} {\bibinfo {title} {Interfacial ferroelectricity by van der {W}aals
  sliding},\ }\href {https://doi.org/10.1126/science.abe8177} {\bibfield
  {journal} {\bibinfo  {journal} {Science}\ }\textbf {\bibinfo {volume}
  {372}},\ \bibinfo {pages} {1462} (\bibinfo {year} {2021})}\BibitemShut
  {NoStop}%
\bibitem [{\citenamefont {Yasuda}\ \emph {et~al.}(2021)\citenamefont {Yasuda},
  \citenamefont {Wang}, \citenamefont {Watanabe}, \citenamefont {Taniguchi},\
  and\ \citenamefont {Jarillo-Herrero}}]{Yasuda2021}%
  \BibitemOpen
  \bibfield  {author} {\bibinfo {author} {\bibfnamefont {K.}~\bibnamefont
  {Yasuda}}, \bibinfo {author} {\bibfnamefont {X.}~\bibnamefont {Wang}},
  \bibinfo {author} {\bibfnamefont {K.}~\bibnamefont {Watanabe}}, \bibinfo
  {author} {\bibfnamefont {T.}~\bibnamefont {Taniguchi}},\ and\ \bibinfo
  {author} {\bibfnamefont {P.}~\bibnamefont {Jarillo-Herrero}},\ }\bibfield
  {title} {\bibinfo {title} {Stacking-engineered ferroelectricity in bilayer
  boron nitride},\ }\href {https://doi.org/10.1126/science.abd3230} {\bibfield
  {journal} {\bibinfo  {journal} {Science}\ }\textbf {\bibinfo {volume}
  {372}},\ \bibinfo {pages} {1458} (\bibinfo {year} {2021})}\BibitemShut
  {NoStop}%
\bibitem [{\citenamefont {Yasuda}\ \emph {et~al.}(2024)\citenamefont {Yasuda},
  \citenamefont {Zalys-Geller}, \citenamefont {Wang}, \citenamefont {Bennett},
  \citenamefont {Cheema}, \citenamefont {Watanabe}, \citenamefont {Taniguchi},
  \citenamefont {Kaxiras}, \citenamefont {Jarillo-Herrero},\ and\ \citenamefont
  {Ashoori}}]{Yasuda2024}%
  \BibitemOpen
  \bibfield  {author} {\bibinfo {author} {\bibfnamefont {K.}~\bibnamefont
  {Yasuda}}, \bibinfo {author} {\bibfnamefont {E.}~\bibnamefont
  {Zalys-Geller}}, \bibinfo {author} {\bibfnamefont {X.}~\bibnamefont {Wang}},
  \bibinfo {author} {\bibfnamefont {D.}~\bibnamefont {Bennett}}, \bibinfo
  {author} {\bibfnamefont {S.~S.}\ \bibnamefont {Cheema}}, \bibinfo {author}
  {\bibfnamefont {K.}~\bibnamefont {Watanabe}}, \bibinfo {author}
  {\bibfnamefont {T.}~\bibnamefont {Taniguchi}}, \bibinfo {author}
  {\bibfnamefont {E.}~\bibnamefont {Kaxiras}}, \bibinfo {author} {\bibfnamefont
  {P.}~\bibnamefont {Jarillo-Herrero}},\ and\ \bibinfo {author} {\bibfnamefont
  {R.}~\bibnamefont {Ashoori}},\ }\bibfield  {title} {\bibinfo {title}
  {Ultrafast high-endurance memory based on sliding ferroelectrics},\ }\href
  {https://doi.org/10.1126/science.adp3575} {\bibfield  {journal} {\bibinfo
  {journal} {Science}\ }\textbf {\bibinfo {volume} {385}},\ \bibinfo {pages}
  {53} (\bibinfo {year} {2024})}\BibitemShut {NoStop}%
\bibitem [{\citenamefont {McGilly}\ \emph {et~al.}(2020)\citenamefont
  {McGilly}, \citenamefont {Kerelsky}, \citenamefont {Finney}, \citenamefont
  {Shapovalov}, \citenamefont {Shih}, \citenamefont {Ghiotto}, \citenamefont
  {Zeng}, \citenamefont {Moore}, \citenamefont {Wu}, \citenamefont {Bai},
  \citenamefont {Watanabe}, \citenamefont {Taniguchi}, \citenamefont {Stengel},
  \citenamefont {Zhou}, \citenamefont {Hone}, \citenamefont {Zhu},
  \citenamefont {Basov}, \citenamefont {Dean}, \citenamefont {Dreyer},\ and\
  \citenamefont {Pasupathy}}]{McGilly2020}%
  \BibitemOpen
  \bibfield  {author} {\bibinfo {author} {\bibfnamefont {L.~J.}\ \bibnamefont
  {McGilly}}, \bibinfo {author} {\bibfnamefont {A.}~\bibnamefont {Kerelsky}},
  \bibinfo {author} {\bibfnamefont {N.~R.}\ \bibnamefont {Finney}}, \bibinfo
  {author} {\bibfnamefont {K.}~\bibnamefont {Shapovalov}}, \bibinfo {author}
  {\bibfnamefont {E.-M.}\ \bibnamefont {Shih}}, \bibinfo {author}
  {\bibfnamefont {A.}~\bibnamefont {Ghiotto}}, \bibinfo {author} {\bibfnamefont
  {Y.}~\bibnamefont {Zeng}}, \bibinfo {author} {\bibfnamefont {S.~L.}\
  \bibnamefont {Moore}}, \bibinfo {author} {\bibfnamefont {W.}~\bibnamefont
  {Wu}}, \bibinfo {author} {\bibfnamefont {Y.}~\bibnamefont {Bai}}, \bibinfo
  {author} {\bibfnamefont {K.}~\bibnamefont {Watanabe}}, \bibinfo {author}
  {\bibfnamefont {T.}~\bibnamefont {Taniguchi}}, \bibinfo {author}
  {\bibfnamefont {M.}~\bibnamefont {Stengel}}, \bibinfo {author} {\bibfnamefont
  {L.}~\bibnamefont {Zhou}}, \bibinfo {author} {\bibfnamefont {J.}~\bibnamefont
  {Hone}}, \bibinfo {author} {\bibfnamefont {X.}~\bibnamefont {Zhu}}, \bibinfo
  {author} {\bibfnamefont {D.~N.}\ \bibnamefont {Basov}}, \bibinfo {author}
  {\bibfnamefont {C.}~\bibnamefont {Dean}}, \bibinfo {author} {\bibfnamefont
  {C.~E.}\ \bibnamefont {Dreyer}},\ and\ \bibinfo {author} {\bibfnamefont
  {A.~N.}\ \bibnamefont {Pasupathy}},\ }\bibfield  {title} {\bibinfo {title}
  {Visualization of moiré superlattices},\ }\href
  {https://doi.org/10.1038/s41565-020-0708-3} {\bibfield  {journal} {\bibinfo
  {journal} {Nat. Nanotechnol.}\ }\textbf {\bibinfo {volume} {15}},\ \bibinfo
  {pages} {580} (\bibinfo {year} {2020})}\BibitemShut {NoStop}%
\bibitem [{\citenamefont {Wang}\ \emph {et~al.}(2022)\citenamefont {Wang},
  \citenamefont {Yasuda}, \citenamefont {Zhang}, \citenamefont {Liu},
  \citenamefont {Watanabe}, \citenamefont {Taniguchi}, \citenamefont {Hone},
  \citenamefont {Fu},\ and\ \citenamefont {Jarillo-Herrero}}]{Wang2022}%
  \BibitemOpen
  \bibfield  {author} {\bibinfo {author} {\bibfnamefont {X.}~\bibnamefont
  {Wang}}, \bibinfo {author} {\bibfnamefont {K.}~\bibnamefont {Yasuda}},
  \bibinfo {author} {\bibfnamefont {Y.}~\bibnamefont {Zhang}}, \bibinfo
  {author} {\bibfnamefont {S.}~\bibnamefont {Liu}}, \bibinfo {author}
  {\bibfnamefont {K.}~\bibnamefont {Watanabe}}, \bibinfo {author}
  {\bibfnamefont {T.}~\bibnamefont {Taniguchi}}, \bibinfo {author}
  {\bibfnamefont {J.}~\bibnamefont {Hone}}, \bibinfo {author} {\bibfnamefont
  {L.}~\bibnamefont {Fu}},\ and\ \bibinfo {author} {\bibfnamefont
  {P.}~\bibnamefont {Jarillo-Herrero}},\ }\bibfield  {title} {\bibinfo {title}
  {Interfacial ferroelectricity in rhombohedral-stacked bilayer transition
  metal dichalcogenides},\ }\href {https://doi.org/10.1038/s41565-021-01059-z}
  {\bibfield  {journal} {\bibinfo  {journal} {Nat. Nanotechnol.}\ }\textbf
  {\bibinfo {volume} {17}},\ \bibinfo {pages} {367} (\bibinfo {year}
  {2022})}\BibitemShut {NoStop}%
\bibitem [{\citenamefont {Weston}\ \emph {et~al.}(2022)\citenamefont {Weston},
  \citenamefont {Castanon}, \citenamefont {Enaldiev}, \citenamefont {Ferreira},
  \citenamefont {Bhattacharjee}, \citenamefont {Xu}, \citenamefont
  {Corte-León}, \citenamefont {Wu}, \citenamefont {Clark}, \citenamefont
  {Summerfield}, \citenamefont {Hashimoto}, \citenamefont {Gao}, \citenamefont
  {Wang}, \citenamefont {Hamer}, \citenamefont {Read}, \citenamefont
  {Fumagalli}, \citenamefont {Kretinin}, \citenamefont {Haigh}, \citenamefont
  {Kazakova}, \citenamefont {Geim}, \citenamefont {Fal’ko},\ and\
  \citenamefont {Gorbachev}}]{Weston2022}%
  \BibitemOpen
  \bibfield  {author} {\bibinfo {author} {\bibfnamefont {A.}~\bibnamefont
  {Weston}}, \bibinfo {author} {\bibfnamefont {E.~G.}\ \bibnamefont
  {Castanon}}, \bibinfo {author} {\bibfnamefont {V.}~\bibnamefont {Enaldiev}},
  \bibinfo {author} {\bibfnamefont {F.}~\bibnamefont {Ferreira}}, \bibinfo
  {author} {\bibfnamefont {S.}~\bibnamefont {Bhattacharjee}}, \bibinfo {author}
  {\bibfnamefont {S.}~\bibnamefont {Xu}}, \bibinfo {author} {\bibfnamefont
  {H.}~\bibnamefont {Corte-León}}, \bibinfo {author} {\bibfnamefont
  {Z.}~\bibnamefont {Wu}}, \bibinfo {author} {\bibfnamefont {N.}~\bibnamefont
  {Clark}}, \bibinfo {author} {\bibfnamefont {A.}~\bibnamefont {Summerfield}},
  \bibinfo {author} {\bibfnamefont {T.}~\bibnamefont {Hashimoto}}, \bibinfo
  {author} {\bibfnamefont {Y.}~\bibnamefont {Gao}}, \bibinfo {author}
  {\bibfnamefont {W.}~\bibnamefont {Wang}}, \bibinfo {author} {\bibfnamefont
  {M.}~\bibnamefont {Hamer}}, \bibinfo {author} {\bibfnamefont
  {H.}~\bibnamefont {Read}}, \bibinfo {author} {\bibfnamefont {L.}~\bibnamefont
  {Fumagalli}}, \bibinfo {author} {\bibfnamefont {A.~V.}\ \bibnamefont
  {Kretinin}}, \bibinfo {author} {\bibfnamefont {S.~J.}\ \bibnamefont {Haigh}},
  \bibinfo {author} {\bibfnamefont {O.}~\bibnamefont {Kazakova}}, \bibinfo
  {author} {\bibfnamefont {A.~K.}\ \bibnamefont {Geim}}, \bibinfo {author}
  {\bibfnamefont {V.~I.}\ \bibnamefont {Fal’ko}},\ and\ \bibinfo {author}
  {\bibfnamefont {R.}~\bibnamefont {Gorbachev}},\ }\bibfield  {title} {\bibinfo
  {title} {Interfacial ferroelectricity in marginally twisted {2D}
  semiconductors},\ }\href {https://doi.org/10.1038/s41565-022-01072-w}
  {\bibfield  {journal} {\bibinfo  {journal} {Nat. Nanotechnol.}\ }\textbf
  {\bibinfo {volume} {17}},\ \bibinfo {pages} {390} (\bibinfo {year}
  {2022})}\BibitemShut {NoStop}%
\bibitem [{\citenamefont {Deb}\ \emph {et~al.}(2022)\citenamefont {Deb},
  \citenamefont {Cao}, \citenamefont {Raab}, \citenamefont {Watanabe},
  \citenamefont {Taniguchi}, \citenamefont {Goldstein}, \citenamefont {Kronik},
  \citenamefont {Urbakh}, \citenamefont {Hod},\ and\ \citenamefont
  {Ben~Shalom}}]{Deb2022}%
  \BibitemOpen
  \bibfield  {author} {\bibinfo {author} {\bibfnamefont {S.}~\bibnamefont
  {Deb}}, \bibinfo {author} {\bibfnamefont {W.}~\bibnamefont {Cao}}, \bibinfo
  {author} {\bibfnamefont {N.}~\bibnamefont {Raab}}, \bibinfo {author}
  {\bibfnamefont {K.}~\bibnamefont {Watanabe}}, \bibinfo {author}
  {\bibfnamefont {T.}~\bibnamefont {Taniguchi}}, \bibinfo {author}
  {\bibfnamefont {M.}~\bibnamefont {Goldstein}}, \bibinfo {author}
  {\bibfnamefont {L.}~\bibnamefont {Kronik}}, \bibinfo {author} {\bibfnamefont
  {M.}~\bibnamefont {Urbakh}}, \bibinfo {author} {\bibfnamefont
  {O.}~\bibnamefont {Hod}},\ and\ \bibinfo {author} {\bibfnamefont
  {M.}~\bibnamefont {Ben~Shalom}},\ }\bibfield  {title} {\bibinfo {title}
  {Cumulative polarization in conductive interfacial ferroelectrics},\ }\href
  {https://doi.org/10.1038/s41586-022-05341-5} {\bibfield  {journal} {\bibinfo
  {journal} {Nature}\ }\textbf {\bibinfo {volume} {612}},\ \bibinfo {pages}
  {465} (\bibinfo {year} {2022})}\BibitemShut {NoStop}%
\bibitem [{\citenamefont {Ko}\ \emph {et~al.}(2023)\citenamefont {Ko},
  \citenamefont {Yuk}, \citenamefont {Engelke}, \citenamefont {Carr},
  \citenamefont {Kim}, \citenamefont {Park}, \citenamefont {Heo}, \citenamefont
  {Kim}, \citenamefont {Kim}, \citenamefont {Kim}, \citenamefont {Taniguchi},
  \citenamefont {Watanabe}, \citenamefont {Park}, \citenamefont {Kaxiras},
  \citenamefont {Yang}, \citenamefont {Kim},\ and\ \citenamefont
  {Yoo}}]{Ko2023}%
  \BibitemOpen
  \bibfield  {author} {\bibinfo {author} {\bibfnamefont {K.}~\bibnamefont
  {Ko}}, \bibinfo {author} {\bibfnamefont {A.}~\bibnamefont {Yuk}}, \bibinfo
  {author} {\bibfnamefont {R.}~\bibnamefont {Engelke}}, \bibinfo {author}
  {\bibfnamefont {S.}~\bibnamefont {Carr}}, \bibinfo {author} {\bibfnamefont
  {J.}~\bibnamefont {Kim}}, \bibinfo {author} {\bibfnamefont {D.}~\bibnamefont
  {Park}}, \bibinfo {author} {\bibfnamefont {H.}~\bibnamefont {Heo}}, \bibinfo
  {author} {\bibfnamefont {H.-M.}\ \bibnamefont {Kim}}, \bibinfo {author}
  {\bibfnamefont {S.-G.}\ \bibnamefont {Kim}}, \bibinfo {author} {\bibfnamefont
  {H.}~\bibnamefont {Kim}}, \bibinfo {author} {\bibfnamefont {T.}~\bibnamefont
  {Taniguchi}}, \bibinfo {author} {\bibfnamefont {K.}~\bibnamefont {Watanabe}},
  \bibinfo {author} {\bibfnamefont {H.}~\bibnamefont {Park}}, \bibinfo {author}
  {\bibfnamefont {E.}~\bibnamefont {Kaxiras}}, \bibinfo {author} {\bibfnamefont
  {S.~M.}\ \bibnamefont {Yang}}, \bibinfo {author} {\bibfnamefont
  {P.}~\bibnamefont {Kim}},\ and\ \bibinfo {author} {\bibfnamefont
  {H.}~\bibnamefont {Yoo}},\ }\bibfield  {title} {\bibinfo {title} {Operando
  electron microscopy investigation of polar domain dynamics in twisted van der
  {W}aals homobilayers},\ }\href {https://doi.org/10.1038/s41563-023-01595-0}
  {\bibfield  {journal} {\bibinfo  {journal} {Nat. Mater.}\ }\textbf {\bibinfo
  {volume} {22}},\ \bibinfo {pages} {992} (\bibinfo {year} {2023})}\BibitemShut
  {NoStop}%
\bibitem [{\citenamefont {Van~Winkle}\ \emph {et~al.}(2024)\citenamefont
  {Van~Winkle}, \citenamefont {Dowlatshahi}, \citenamefont {Khaloo},
  \citenamefont {Iyer}, \citenamefont {Craig}, \citenamefont {Dhall},
  \citenamefont {Taniguchi}, \citenamefont {Watanabe},\ and\ \citenamefont
  {Bediako}}]{Winkle2024}%
  \BibitemOpen
  \bibfield  {author} {\bibinfo {author} {\bibfnamefont {M.}~\bibnamefont
  {Van~Winkle}}, \bibinfo {author} {\bibfnamefont {N.}~\bibnamefont
  {Dowlatshahi}}, \bibinfo {author} {\bibfnamefont {N.}~\bibnamefont {Khaloo}},
  \bibinfo {author} {\bibfnamefont {M.}~\bibnamefont {Iyer}}, \bibinfo {author}
  {\bibfnamefont {I.~M.}\ \bibnamefont {Craig}}, \bibinfo {author}
  {\bibfnamefont {R.}~\bibnamefont {Dhall}}, \bibinfo {author} {\bibfnamefont
  {T.}~\bibnamefont {Taniguchi}}, \bibinfo {author} {\bibfnamefont
  {K.}~\bibnamefont {Watanabe}},\ and\ \bibinfo {author} {\bibfnamefont
  {D.~K.}\ \bibnamefont {Bediako}},\ }\bibfield  {title} {\bibinfo {title}
  {Engineering interfacial polarization switching in van der {W}aals
  multilayers},\ }\href {https://doi.org/10.1038/s41565-024-01642-0} {\bibfield
   {journal} {\bibinfo  {journal} {Nat. Nanotechnol.}\ }\textbf {\bibinfo
  {volume} {19}},\ \bibinfo {pages} {751} (\bibinfo {year} {2024})}\BibitemShut
  {NoStop}%
\bibitem [{\citenamefont {Zheng}\ \emph {et~al.}(2020)\citenamefont {Zheng},
  \citenamefont {Ma}, \citenamefont {Bi}, \citenamefont {De~La~Barrera},
  \citenamefont {Liu}, \citenamefont {Mao}, \citenamefont {Zhang},
  \citenamefont {Kiper}, \citenamefont {Watanabe}, \citenamefont {Taniguchi},
  \citenamefont {Kong}, \citenamefont {Tisdale}, \citenamefont {Ashoori},
  \citenamefont {Gedik}, \citenamefont {Fu}, \citenamefont {Xu},\ and\
  \citenamefont {Jarillo-Herrero}}]{Zheng2020}%
  \BibitemOpen
  \bibfield  {author} {\bibinfo {author} {\bibfnamefont {Z.}~\bibnamefont
  {Zheng}}, \bibinfo {author} {\bibfnamefont {Q.}~\bibnamefont {Ma}}, \bibinfo
  {author} {\bibfnamefont {Z.}~\bibnamefont {Bi}}, \bibinfo {author}
  {\bibfnamefont {S.}~\bibnamefont {De~La~Barrera}}, \bibinfo {author}
  {\bibfnamefont {M.-H.}\ \bibnamefont {Liu}}, \bibinfo {author} {\bibfnamefont
  {N.}~\bibnamefont {Mao}}, \bibinfo {author} {\bibfnamefont {Y.}~\bibnamefont
  {Zhang}}, \bibinfo {author} {\bibfnamefont {N.}~\bibnamefont {Kiper}},
  \bibinfo {author} {\bibfnamefont {K.}~\bibnamefont {Watanabe}}, \bibinfo
  {author} {\bibfnamefont {T.}~\bibnamefont {Taniguchi}}, \bibinfo {author}
  {\bibfnamefont {J.}~\bibnamefont {Kong}}, \bibinfo {author} {\bibfnamefont
  {W.~A.}\ \bibnamefont {Tisdale}}, \bibinfo {author} {\bibfnamefont
  {R.}~\bibnamefont {Ashoori}}, \bibinfo {author} {\bibfnamefont
  {N.}~\bibnamefont {Gedik}}, \bibinfo {author} {\bibfnamefont
  {L.}~\bibnamefont {Fu}}, \bibinfo {author} {\bibfnamefont {S.-Y.}\
  \bibnamefont {Xu}},\ and\ \bibinfo {author} {\bibfnamefont {P.}~\bibnamefont
  {Jarillo-Herrero}},\ }\bibfield  {title} {\bibinfo {title} {Unconventional
  ferroelectricity in moiré heterostructures},\ }\href
  {https://doi.org/10.1038/s41586-020-2970-9} {\bibfield  {journal} {\bibinfo
  {journal} {Nature}\ }\textbf {\bibinfo {volume} {588}},\ \bibinfo {pages}
  {71} (\bibinfo {year} {2020})}\BibitemShut {NoStop}%
\bibitem [{\citenamefont {Li}\ \emph {et~al.}(2024)\citenamefont {Li},
  \citenamefont {Qin}, \citenamefont {Wang}, \citenamefont {Xi}, \citenamefont
  {Huang}, \citenamefont {Zhao}, \citenamefont {Peng}, \citenamefont {Chen},
  \citenamefont {Pan}, \citenamefont {Zhu}, \citenamefont {Cui}, \citenamefont
  {Yang}, \citenamefont {Yang}, \citenamefont {Meng}, \citenamefont {Shi},
  \citenamefont {Bai}, \citenamefont {Liu}, \citenamefont {Li}, \citenamefont
  {Tang}, \citenamefont {Liu}, \citenamefont {Du},\ and\ \citenamefont
  {Zhang}}]{Li2024}%
  \BibitemOpen
  \bibfield  {author} {\bibinfo {author} {\bibfnamefont {X.}~\bibnamefont
  {Li}}, \bibinfo {author} {\bibfnamefont {B.}~\bibnamefont {Qin}}, \bibinfo
  {author} {\bibfnamefont {Y.}~\bibnamefont {Wang}}, \bibinfo {author}
  {\bibfnamefont {Y.}~\bibnamefont {Xi}}, \bibinfo {author} {\bibfnamefont
  {Z.}~\bibnamefont {Huang}}, \bibinfo {author} {\bibfnamefont
  {M.}~\bibnamefont {Zhao}}, \bibinfo {author} {\bibfnamefont {Y.}~\bibnamefont
  {Peng}}, \bibinfo {author} {\bibfnamefont {Z.}~\bibnamefont {Chen}}, \bibinfo
  {author} {\bibfnamefont {Z.}~\bibnamefont {Pan}}, \bibinfo {author}
  {\bibfnamefont {J.}~\bibnamefont {Zhu}}, \bibinfo {author} {\bibfnamefont
  {C.}~\bibnamefont {Cui}}, \bibinfo {author} {\bibfnamefont {R.}~\bibnamefont
  {Yang}}, \bibinfo {author} {\bibfnamefont {W.}~\bibnamefont {Yang}}, \bibinfo
  {author} {\bibfnamefont {S.}~\bibnamefont {Meng}}, \bibinfo {author}
  {\bibfnamefont {D.}~\bibnamefont {Shi}}, \bibinfo {author} {\bibfnamefont
  {X.}~\bibnamefont {Bai}}, \bibinfo {author} {\bibfnamefont {C.}~\bibnamefont
  {Liu}}, \bibinfo {author} {\bibfnamefont {N.}~\bibnamefont {Li}}, \bibinfo
  {author} {\bibfnamefont {J.}~\bibnamefont {Tang}}, \bibinfo {author}
  {\bibfnamefont {K.}~\bibnamefont {Liu}}, \bibinfo {author} {\bibfnamefont
  {L.}~\bibnamefont {Du}},\ and\ \bibinfo {author} {\bibfnamefont
  {G.}~\bibnamefont {Zhang}},\ }\bibfield  {title} {\bibinfo {title} {Sliding
  ferroelectric memories and synapses based on rhombohedral-stacked bilayer
  {MoS}$_{\textrm{2}}$},\ }\href {https://doi.org/10.1038/s41467-024-55333-4}
  {\bibfield  {journal} {\bibinfo  {journal} {Nat. Commun.}\ }\textbf {\bibinfo
  {volume} {15}},\ \bibinfo {pages} {10921} (\bibinfo {year}
  {2024})}\BibitemShut {NoStop}%
\bibitem [{\citenamefont {Bian}\ \emph {et~al.}(2024)\citenamefont {Bian},
  \citenamefont {He}, \citenamefont {Pan}, \citenamefont {Li}, \citenamefont
  {Cao}, \citenamefont {Meng}, \citenamefont {Chen}, \citenamefont {Liu},
  \citenamefont {Zhong}, \citenamefont {Li},\ and\ \citenamefont
  {Liu}}]{Bian2024}%
  \BibitemOpen
  \bibfield  {author} {\bibinfo {author} {\bibfnamefont {R.}~\bibnamefont
  {Bian}}, \bibinfo {author} {\bibfnamefont {R.}~\bibnamefont {He}}, \bibinfo
  {author} {\bibfnamefont {E.}~\bibnamefont {Pan}}, \bibinfo {author}
  {\bibfnamefont {Z.}~\bibnamefont {Li}}, \bibinfo {author} {\bibfnamefont
  {G.}~\bibnamefont {Cao}}, \bibinfo {author} {\bibfnamefont {P.}~\bibnamefont
  {Meng}}, \bibinfo {author} {\bibfnamefont {J.}~\bibnamefont {Chen}}, \bibinfo
  {author} {\bibfnamefont {Q.}~\bibnamefont {Liu}}, \bibinfo {author}
  {\bibfnamefont {Z.}~\bibnamefont {Zhong}}, \bibinfo {author} {\bibfnamefont
  {W.}~\bibnamefont {Li}},\ and\ \bibinfo {author} {\bibfnamefont
  {F.}~\bibnamefont {Liu}},\ }\bibfield  {title} {\bibinfo {title} {Developing
  fatigue-resistant ferroelectrics using interlayer sliding switching},\ }\href
  {https://doi.org/10.1126/science.ado1744} {\bibfield  {journal} {\bibinfo
  {journal} {Science}\ }\textbf {\bibinfo {volume} {385}},\ \bibinfo {pages}
  {57} (\bibinfo {year} {2024})}\BibitemShut {NoStop}%
\bibitem [{\citenamefont {Niu}\ \emph {et~al.}(2022)\citenamefont {Niu},
  \citenamefont {Li}, \citenamefont {Han}, \citenamefont {Qu}, \citenamefont
  {Ding}, \citenamefont {Wang}, \citenamefont {Liu}, \citenamefont {Liu},
  \citenamefont {Han}, \citenamefont {Watanabe}, \citenamefont {Taniguchi},
  \citenamefont {Wu}, \citenamefont {Ren}, \citenamefont {Wang}, \citenamefont
  {Hong}, \citenamefont {Mao}, \citenamefont {Han}, \citenamefont {Liu},
  \citenamefont {Gan},\ and\ \citenamefont {Lu}}]{Niu2022}%
  \BibitemOpen
  \bibfield  {author} {\bibinfo {author} {\bibfnamefont {R.}~\bibnamefont
  {Niu}}, \bibinfo {author} {\bibfnamefont {Z.}~\bibnamefont {Li}}, \bibinfo
  {author} {\bibfnamefont {X.}~\bibnamefont {Han}}, \bibinfo {author}
  {\bibfnamefont {Z.}~\bibnamefont {Qu}}, \bibinfo {author} {\bibfnamefont
  {D.}~\bibnamefont {Ding}}, \bibinfo {author} {\bibfnamefont {Z.}~\bibnamefont
  {Wang}}, \bibinfo {author} {\bibfnamefont {Q.}~\bibnamefont {Liu}}, \bibinfo
  {author} {\bibfnamefont {T.}~\bibnamefont {Liu}}, \bibinfo {author}
  {\bibfnamefont {C.}~\bibnamefont {Han}}, \bibinfo {author} {\bibfnamefont
  {K.}~\bibnamefont {Watanabe}}, \bibinfo {author} {\bibfnamefont
  {T.}~\bibnamefont {Taniguchi}}, \bibinfo {author} {\bibfnamefont
  {M.}~\bibnamefont {Wu}}, \bibinfo {author} {\bibfnamefont {Q.}~\bibnamefont
  {Ren}}, \bibinfo {author} {\bibfnamefont {X.}~\bibnamefont {Wang}}, \bibinfo
  {author} {\bibfnamefont {J.}~\bibnamefont {Hong}}, \bibinfo {author}
  {\bibfnamefont {J.}~\bibnamefont {Mao}}, \bibinfo {author} {\bibfnamefont
  {Z.}~\bibnamefont {Han}}, \bibinfo {author} {\bibfnamefont {K.}~\bibnamefont
  {Liu}}, \bibinfo {author} {\bibfnamefont {Z.}~\bibnamefont {Gan}},\ and\
  \bibinfo {author} {\bibfnamefont {J.}~\bibnamefont {Lu}},\ }\bibfield
  {title} {\bibinfo {title} {Giant ferroelectric polarization in a bilayer
  graphene heterostructure},\ }\href
  {https://doi.org/10.1038/s41467-022-34104-z} {\bibfield  {journal} {\bibinfo
  {journal} {Nat. Commun.}\ }\textbf {\bibinfo {volume} {13}},\ \bibinfo
  {pages} {6241} (\bibinfo {year} {2022})}\BibitemShut {NoStop}%
\bibitem [{\citenamefont {Klein}\ \emph {et~al.}(2023)\citenamefont {Klein},
  \citenamefont {Xia}, \citenamefont {MacNeill}, \citenamefont {Watanabe},
  \citenamefont {Taniguchi},\ and\ \citenamefont
  {Jarillo-Herrero}}]{Klein2023}%
  \BibitemOpen
  \bibfield  {author} {\bibinfo {author} {\bibfnamefont {D.~R.}\ \bibnamefont
  {Klein}}, \bibinfo {author} {\bibfnamefont {L.-Q.}\ \bibnamefont {Xia}},
  \bibinfo {author} {\bibfnamefont {D.}~\bibnamefont {MacNeill}}, \bibinfo
  {author} {\bibfnamefont {K.}~\bibnamefont {Watanabe}}, \bibinfo {author}
  {\bibfnamefont {T.}~\bibnamefont {Taniguchi}},\ and\ \bibinfo {author}
  {\bibfnamefont {P.}~\bibnamefont {Jarillo-Herrero}},\ }\bibfield  {title}
  {\bibinfo {title} {Electrical switching of a bistable moiré
  superconductor},\ }\href {https://doi.org/10.1038/s41565-022-01314-x}
  {\bibfield  {journal} {\bibinfo  {journal} {Nat. Nanotechnol.}\ }\textbf
  {\bibinfo {volume} {18}},\ \bibinfo {pages} {331} (\bibinfo {year}
  {2023})}\BibitemShut {NoStop}%
\bibitem [{\citenamefont {Chen}\ \emph {et~al.}(2024)\citenamefont {Chen},
  \citenamefont {Xie}, \citenamefont {Cheng}, \citenamefont {Yang},
  \citenamefont {Li}, \citenamefont {Chen}, \citenamefont {Li}, \citenamefont
  {Xie}, \citenamefont {Watanabe}, \citenamefont {Taniguchi}, \citenamefont
  {He}, \citenamefont {Wu}, \citenamefont {Liang},\ and\ \citenamefont
  {Miao}}]{Chen2024}%
  \BibitemOpen
  \bibfield  {author} {\bibinfo {author} {\bibfnamefont {M.}~\bibnamefont
  {Chen}}, \bibinfo {author} {\bibfnamefont {Y.}~\bibnamefont {Xie}}, \bibinfo
  {author} {\bibfnamefont {B.}~\bibnamefont {Cheng}}, \bibinfo {author}
  {\bibfnamefont {Z.}~\bibnamefont {Yang}}, \bibinfo {author} {\bibfnamefont
  {X.-Z.}\ \bibnamefont {Li}}, \bibinfo {author} {\bibfnamefont
  {F.}~\bibnamefont {Chen}}, \bibinfo {author} {\bibfnamefont {Q.}~\bibnamefont
  {Li}}, \bibinfo {author} {\bibfnamefont {J.}~\bibnamefont {Xie}}, \bibinfo
  {author} {\bibfnamefont {K.}~\bibnamefont {Watanabe}}, \bibinfo {author}
  {\bibfnamefont {T.}~\bibnamefont {Taniguchi}}, \bibinfo {author}
  {\bibfnamefont {W.-Y.}\ \bibnamefont {He}}, \bibinfo {author} {\bibfnamefont
  {M.}~\bibnamefont {Wu}}, \bibinfo {author} {\bibfnamefont {S.-J.}\
  \bibnamefont {Liang}},\ and\ \bibinfo {author} {\bibfnamefont
  {F.}~\bibnamefont {Miao}},\ }\bibfield  {title} {\bibinfo {title} {Selective
  and quasi-continuous switching of ferroelectric {C}hern insulator devices for
  neuromorphic computing},\ }\href {https://doi.org/10.1038/s41565-024-01698-y}
  {\bibfield  {journal} {\bibinfo  {journal} {Nat. Nanotechnol.}\ }\textbf
  {\bibinfo {volume} {19}},\ \bibinfo {pages} {962} (\bibinfo {year}
  {2024})}\BibitemShut {NoStop}%
\bibitem [{\citenamefont {He}\ \emph {et~al.}(2024)\citenamefont {He},
  \citenamefont {Zhang}, \citenamefont {Wang}, \citenamefont {Li},
  \citenamefont {Tang}, \citenamefont {Bauer},\ and\ \citenamefont
  {Zhong}}]{He2024}%
  \BibitemOpen
  \bibfield  {author} {\bibinfo {author} {\bibfnamefont {R.}~\bibnamefont
  {He}}, \bibinfo {author} {\bibfnamefont {B.}~\bibnamefont {Zhang}}, \bibinfo
  {author} {\bibfnamefont {H.}~\bibnamefont {Wang}}, \bibinfo {author}
  {\bibfnamefont {L.}~\bibnamefont {Li}}, \bibinfo {author} {\bibfnamefont
  {P.}~\bibnamefont {Tang}}, \bibinfo {author} {\bibfnamefont {G.}~\bibnamefont
  {Bauer}},\ and\ \bibinfo {author} {\bibfnamefont {Z.}~\bibnamefont {Zhong}},\
  }\bibfield  {title} {\bibinfo {title} {Ultrafast switching dynamics of the
  ferroelectric order in stacking-engineered ferroelectrics},\ }\href
  {https://doi.org/https://doi.org/10.1016/j.actamat.2023.119416} {\bibfield
  {journal} {\bibinfo  {journal} {Acta Mater.}\ }\textbf {\bibinfo {volume}
  {262}},\ \bibinfo {pages} {119416} (\bibinfo {year} {2024})}\BibitemShut
  {NoStop}%
\bibitem [{\citenamefont {Wang}\ and\ \citenamefont {Dong}(2025)}]{Wang2025}%
  \BibitemOpen
  \bibfield  {author} {\bibinfo {author} {\bibfnamefont {Z.}~\bibnamefont
  {Wang}}\ and\ \bibinfo {author} {\bibfnamefont {S.}~\bibnamefont {Dong}},\
  }\bibfield  {title} {\bibinfo {title} {Polarization switching in sliding
  ferroelectrics: Roles of fluctuation and domain wall},\ }\href
  {https://doi.org/10.1103/PhysRevB.111.L201406} {\bibfield  {journal}
  {\bibinfo  {journal} {Phys. Rev. B}\ }\textbf {\bibinfo {volume} {111}},\
  \bibinfo {pages} {L201406} (\bibinfo {year} {2025})}\BibitemShut {NoStop}%
\bibitem [{\citenamefont {Ke}\ \emph {et~al.}(2025)\citenamefont {Ke},
  \citenamefont {Liu},\ and\ \citenamefont {Liu}}]{Ke2025}%
  \BibitemOpen
  \bibfield  {author} {\bibinfo {author} {\bibfnamefont {C.}~\bibnamefont
  {Ke}}, \bibinfo {author} {\bibfnamefont {F.}~\bibnamefont {Liu}},\ and\
  \bibinfo {author} {\bibfnamefont {S.}~\bibnamefont {Liu}},\ }\bibfield
  {title} {\bibinfo {title} {Superlubric motion of wavelike domain walls in
  sliding ferroelectrics},\ }\href {https://doi.org/10.1103/jhlq-2dd7}
  {\bibfield  {journal} {\bibinfo  {journal} {Phys. Rev. Lett.}\ }\textbf
  {\bibinfo {volume} {135}},\ \bibinfo {pages} {046201} (\bibinfo {year}
  {2025})}\BibitemShut {NoStop}%
\bibitem [{\citenamefont {Shi}\ \emph {et~al.}(2025)\citenamefont {Shi},
  \citenamefont {Gao}, \citenamefont {Wang}, \citenamefont {Zhang},
  \citenamefont {Zhong},\ and\ \citenamefont {He}}]{Shi2025}%
  \BibitemOpen
  \bibfield  {author} {\bibinfo {author} {\bibfnamefont {Y.}~\bibnamefont
  {Shi}}, \bibinfo {author} {\bibfnamefont {Y.}~\bibnamefont {Gao}}, \bibinfo
  {author} {\bibfnamefont {H.}~\bibnamefont {Wang}}, \bibinfo {author}
  {\bibfnamefont {B.}~\bibnamefont {Zhang}}, \bibinfo {author} {\bibfnamefont
  {Z.}~\bibnamefont {Zhong}},\ and\ \bibinfo {author} {\bibfnamefont
  {R.}~\bibnamefont {He}},\ }\bibfield  {title} {\bibinfo {title} {Soliton-like
  domain wall motion in sliding ferroelectrics with ultralow damping},\ }\href
  {https://doi.org/10.1103/b91v-r2rc} {\bibfield  {journal} {\bibinfo
  {journal} {Phys. Rev. B}\ }\textbf {\bibinfo {volume} {112}},\ \bibinfo
  {pages} {035421} (\bibinfo {year} {2025})}\BibitemShut {NoStop}%
\bibitem [{\citenamefont {Bai}\ \emph {et~al.}(2025)\citenamefont {Bai},
  \citenamefont {Yu}, \citenamefont {Guan}, \citenamefont {Tian}, \citenamefont
  {Wang}, \citenamefont {Yao}, \citenamefont {Yang}, \citenamefont {Lei},
  \citenamefont {Xu}, \citenamefont {Liu}, \citenamefont {Zhu}, \citenamefont
  {Tu}, \citenamefont {Shen}, \citenamefont {Xiang}, \citenamefont {Li},
  \citenamefont {Xu},\ and\ \citenamefont {Wang}}]{Bai2025}%
  \BibitemOpen
  \bibfield  {author} {\bibinfo {author} {\bibfnamefont {Y.}~\bibnamefont
  {Bai}}, \bibinfo {author} {\bibfnamefont {Z.}~\bibnamefont {Yu}}, \bibinfo
  {author} {\bibfnamefont {Z.}~\bibnamefont {Guan}}, \bibinfo {author}
  {\bibfnamefont {J.}~\bibnamefont {Tian}}, \bibinfo {author} {\bibfnamefont
  {C.}~\bibnamefont {Wang}}, \bibinfo {author} {\bibfnamefont {X.}~\bibnamefont
  {Yao}}, \bibinfo {author} {\bibfnamefont {Y.}~\bibnamefont {Yang}}, \bibinfo
  {author} {\bibfnamefont {Y.}~\bibnamefont {Lei}}, \bibinfo {author}
  {\bibfnamefont {J.}~\bibnamefont {Xu}}, \bibinfo {author} {\bibfnamefont
  {C.}~\bibnamefont {Liu}}, \bibinfo {author} {\bibfnamefont {J.}~\bibnamefont
  {Zhu}}, \bibinfo {author} {\bibfnamefont {Y.}~\bibnamefont {Tu}}, \bibinfo
  {author} {\bibfnamefont {S.}~\bibnamefont {Shen}}, \bibinfo {author}
  {\bibfnamefont {H.}~\bibnamefont {Xiang}}, \bibinfo {author} {\bibfnamefont
  {X.}~\bibnamefont {Li}}, \bibinfo {author} {\bibfnamefont {C.}~\bibnamefont
  {Xu}},\ and\ \bibinfo {author} {\bibfnamefont {J.}~\bibnamefont {Wang}},\
  }\bibfield  {title} {\bibinfo {title} {Sub-nanosecond polarization switching
  with anomalous kinetics in {vdW} ferroelectric {WTe$_2$}},\ }\href
  {https://doi.org/10.1038/s41467-025-62608-x} {\bibfield  {journal} {\bibinfo
  {journal} {Nat. Commun.}\ }\textbf {\bibinfo {volume} {16}},\ \bibinfo
  {pages} {7221} (\bibinfo {year} {2025})}\BibitemShut {NoStop}%
\bibitem [{\citenamefont {Liu}\ \emph {et~al.}(2025)\citenamefont {Liu},
  \citenamefont {Li}, \citenamefont {Gong}, \citenamefont {Wen}, \citenamefont
  {Zhou}, \citenamefont {Feng}, \citenamefont {Zhang}, \citenamefont {Zou},
  \citenamefont {Wu}, \citenamefont {Li}, \citenamefont {Zhu}, \citenamefont
  {Zhuo}, \citenamefont {Zou}, \citenamefont {Hu}, \citenamefont {Ding},
  \citenamefont {Fang}, \citenamefont {Xu}, \citenamefont {Hou}, \citenamefont
  {Zhang}, \citenamefont {Long}, \citenamefont {Tang}, \citenamefont {Jiang},
  \citenamefont {Yu}, \citenamefont {Ma}, \citenamefont {Wang},\ and\
  \citenamefont {Wang}}]{Liu2025b}%
  \BibitemOpen
  \bibfield  {author} {\bibinfo {author} {\bibfnamefont {L.}~\bibnamefont
  {Liu}}, \bibinfo {author} {\bibfnamefont {T.}~\bibnamefont {Li}}, \bibinfo
  {author} {\bibfnamefont {X.}~\bibnamefont {Gong}}, \bibinfo {author}
  {\bibfnamefont {H.}~\bibnamefont {Wen}}, \bibinfo {author} {\bibfnamefont
  {L.}~\bibnamefont {Zhou}}, \bibinfo {author} {\bibfnamefont {M.}~\bibnamefont
  {Feng}}, \bibinfo {author} {\bibfnamefont {H.}~\bibnamefont {Zhang}},
  \bibinfo {author} {\bibfnamefont {N.}~\bibnamefont {Zou}}, \bibinfo {author}
  {\bibfnamefont {S.}~\bibnamefont {Wu}}, \bibinfo {author} {\bibfnamefont
  {Y.}~\bibnamefont {Li}}, \bibinfo {author} {\bibfnamefont {S.}~\bibnamefont
  {Zhu}}, \bibinfo {author} {\bibfnamefont {F.}~\bibnamefont {Zhuo}}, \bibinfo
  {author} {\bibfnamefont {X.}~\bibnamefont {Zou}}, \bibinfo {author}
  {\bibfnamefont {Z.}~\bibnamefont {Hu}}, \bibinfo {author} {\bibfnamefont
  {Z.}~\bibnamefont {Ding}}, \bibinfo {author} {\bibfnamefont {S.}~\bibnamefont
  {Fang}}, \bibinfo {author} {\bibfnamefont {W.}~\bibnamefont {Xu}}, \bibinfo
  {author} {\bibfnamefont {X.}~\bibnamefont {Hou}}, \bibinfo {author}
  {\bibfnamefont {K.}~\bibnamefont {Zhang}}, \bibinfo {author} {\bibfnamefont
  {G.}~\bibnamefont {Long}}, \bibinfo {author} {\bibfnamefont {L.}~\bibnamefont
  {Tang}}, \bibinfo {author} {\bibfnamefont {Y.}~\bibnamefont {Jiang}},
  \bibinfo {author} {\bibfnamefont {Z.}~\bibnamefont {Yu}}, \bibinfo {author}
  {\bibfnamefont {L.}~\bibnamefont {Ma}}, \bibinfo {author} {\bibfnamefont
  {J.}~\bibnamefont {Wang}},\ and\ \bibinfo {author} {\bibfnamefont
  {X.}~\bibnamefont {Wang}},\ }\bibfield  {title} {\bibinfo {title}
  {Homoepitaxial growth of large-area rhombohedral-stacked {MoS}$_2$},\ }\href
  {https://doi.org/10.1038/s41563-025-02274-y} {\bibfield  {journal} {\bibinfo
  {journal} {Nat. Mater.}\ }\textbf {\bibinfo {volume} {24}},\ \bibinfo {pages}
  {1195} (\bibinfo {year} {2025})}\BibitemShut {NoStop}%
\bibitem [{\citenamefont {Lu}\ \emph {et~al.}(2015)\citenamefont {Lu},
  \citenamefont {Utama}, \citenamefont {Lin}, \citenamefont {Luo},
  \citenamefont {Zhao}, \citenamefont {Zhang}, \citenamefont {Pantelides},
  \citenamefont {Zhou}, \citenamefont {Quek},\ and\ \citenamefont
  {Xiong}}]{Lu2015}%
  \BibitemOpen
  \bibfield  {author} {\bibinfo {author} {\bibfnamefont {X.}~\bibnamefont
  {Lu}}, \bibinfo {author} {\bibfnamefont {M.~I.~B.}\ \bibnamefont {Utama}},
  \bibinfo {author} {\bibfnamefont {J.}~\bibnamefont {Lin}}, \bibinfo {author}
  {\bibfnamefont {X.}~\bibnamefont {Luo}}, \bibinfo {author} {\bibfnamefont
  {Y.}~\bibnamefont {Zhao}}, \bibinfo {author} {\bibfnamefont {J.}~\bibnamefont
  {Zhang}}, \bibinfo {author} {\bibfnamefont {S.~T.}\ \bibnamefont
  {Pantelides}}, \bibinfo {author} {\bibfnamefont {W.}~\bibnamefont {Zhou}},
  \bibinfo {author} {\bibfnamefont {S.~Y.}\ \bibnamefont {Quek}},\ and\
  \bibinfo {author} {\bibfnamefont {Q.}~\bibnamefont {Xiong}},\ }\bibfield
  {title} {\bibinfo {title} {Rapid and nondestructive identification of
  polytypism and stacking sequences in few-layer molybdenum diselenide by raman
  spectroscopy},\ }\href
  {https://doi.org/https://doi.org/10.1002/adma.201501086} {\bibfield
  {journal} {\bibinfo  {journal} {Advanced Materials}\ }\textbf {\bibinfo
  {volume} {27}},\ \bibinfo {pages} {4502} (\bibinfo {year}
  {2015})}\BibitemShut {NoStop}%
\bibitem [{\citenamefont {Luo}\ \emph {et~al.}(2015)\citenamefont {Luo},
  \citenamefont {Lu}, \citenamefont {Cong}, \citenamefont {Yu}, \citenamefont
  {Xiong},\ and\ \citenamefont {Ying~Quek}}]{Luo2015}%
  \BibitemOpen
  \bibfield  {author} {\bibinfo {author} {\bibfnamefont {X.}~\bibnamefont
  {Luo}}, \bibinfo {author} {\bibfnamefont {X.}~\bibnamefont {Lu}}, \bibinfo
  {author} {\bibfnamefont {C.}~\bibnamefont {Cong}}, \bibinfo {author}
  {\bibfnamefont {T.}~\bibnamefont {Yu}}, \bibinfo {author} {\bibfnamefont
  {Q.}~\bibnamefont {Xiong}},\ and\ \bibinfo {author} {\bibfnamefont
  {S.}~\bibnamefont {Ying~Quek}},\ }\bibfield  {title} {\bibinfo {title}
  {Stacking sequence determines {R}aman intensities of observed interlayer
  shear modes in {2D} layered materials -- {A} general bond polarizability
  model},\ }\href {https://doi.org/10.1038/srep14565} {\bibfield  {journal}
  {\bibinfo  {journal} {Sci. Rep.}\ }\textbf {\bibinfo {volume} {5}},\ \bibinfo
  {pages} {14565} (\bibinfo {year} {2015})}\BibitemShut {NoStop}%
\bibitem [{\citenamefont {Lee}\ \emph {et~al.}(2016)\citenamefont {Lee},
  \citenamefont {Kim}, \citenamefont {Han}, \citenamefont {Ryu}, \citenamefont
  {Lee},\ and\ \citenamefont {Cheong}}]{Lee2016}%
  \BibitemOpen
  \bibfield  {author} {\bibinfo {author} {\bibfnamefont {J.-U.}\ \bibnamefont
  {Lee}}, \bibinfo {author} {\bibfnamefont {K.}~\bibnamefont {Kim}}, \bibinfo
  {author} {\bibfnamefont {S.}~\bibnamefont {Han}}, \bibinfo {author}
  {\bibfnamefont {G.~H.}\ \bibnamefont {Ryu}}, \bibinfo {author} {\bibfnamefont
  {Z.}~\bibnamefont {Lee}},\ and\ \bibinfo {author} {\bibfnamefont
  {H.}~\bibnamefont {Cheong}},\ }\bibfield  {title} {\bibinfo {title} {Raman
  signatures of polytypism in molybdenum disulfide},\ }\href
  {https://doi.org/10.1021/acsnano.5b05831} {\bibfield  {journal} {\bibinfo
  {journal} {ACS Nano}\ }\textbf {\bibinfo {volume} {10}},\ \bibinfo {pages}
  {1948} (\bibinfo {year} {2016})}\BibitemShut {NoStop}%
\bibitem [{\citenamefont {Liang}\ \emph {et~al.}(2017)\citenamefont {Liang},
  \citenamefont {Puretzky}, \citenamefont {Sumpter},\ and\ \citenamefont
  {Meunier}}]{Liang2017}%
  \BibitemOpen
  \bibfield  {author} {\bibinfo {author} {\bibfnamefont {L.}~\bibnamefont
  {Liang}}, \bibinfo {author} {\bibfnamefont {A.~A.}\ \bibnamefont {Puretzky}},
  \bibinfo {author} {\bibfnamefont {B.~G.}\ \bibnamefont {Sumpter}},\ and\
  \bibinfo {author} {\bibfnamefont {V.}~\bibnamefont {Meunier}},\ }\bibfield
  {title} {\bibinfo {title} {Interlayer bond polarizability model for
  stacking-dependent low-frequency raman scattering in layered materials},\
  }\href {https://doi.org/10.1039/C7NR05839J} {\bibfield  {journal} {\bibinfo
  {journal} {Nanoscale}\ }\textbf {\bibinfo {volume} {9}},\ \bibinfo {pages}
  {15340} (\bibinfo {year} {2017})}\BibitemShut {NoStop}%
\bibitem [{\citenamefont {Lin}\ and\ \citenamefont {Tan}(2019)}]{Lin2019}%
  \BibitemOpen
  \bibfield  {author} {\bibinfo {author} {\bibfnamefont {M.-L.}\ \bibnamefont
  {Lin}}\ and\ \bibinfo {author} {\bibfnamefont {P.-H.}\ \bibnamefont {Tan}},\
  }\bibinfo {title} {Ultralow-frequency raman spectroscopy of two-dimensional
  materials},\ in\ \href {https://doi.org/10.1007/978-981-13-1828-3_10} {\emph
  {\bibinfo {booktitle} {Raman Spectroscopy of Two-Dimensional Materials}}},\
  \bibinfo {editor} {edited by\ \bibinfo {editor} {\bibfnamefont {P.-H.}\
  \bibnamefont {Tan}}}\ (\bibinfo  {publisher} {Springer Singapore},\ \bibinfo
  {address} {Singapore},\ \bibinfo {year} {2019})\ pp.\ \bibinfo {pages}
  {203--230}\BibitemShut {NoStop}%
\bibitem [{\citenamefont {Cong}\ \emph {et~al.}(2020)\citenamefont {Cong},
  \citenamefont {Liu}, \citenamefont {Lin},\ and\ \citenamefont
  {Tan}}]{Cong2020}%
  \BibitemOpen
  \bibfield  {author} {\bibinfo {author} {\bibfnamefont {X.}~\bibnamefont
  {Cong}}, \bibinfo {author} {\bibfnamefont {X.-L.}\ \bibnamefont {Liu}},
  \bibinfo {author} {\bibfnamefont {M.-L.}\ \bibnamefont {Lin}},\ and\ \bibinfo
  {author} {\bibfnamefont {P.-H.}\ \bibnamefont {Tan}},\ }\bibfield  {title}
  {\bibinfo {title} {Application of {R}aman spectroscopy to probe fundamental
  properties of two-dimensional materials},\ }\href
  {https://doi.org/10.1038/s41699-020-0140-4} {\bibfield  {journal} {\bibinfo
  {journal} {npj 2D Mater. Appl.}\ }\textbf {\bibinfo {volume} {4}},\ \bibinfo
  {pages} {13} (\bibinfo {year} {2020})}\BibitemShut {NoStop}%
\bibitem [{\citenamefont {van Baren}\ \emph {et~al.}(2019)\citenamefont {van
  Baren}, \citenamefont {Ye}, \citenamefont {Yan}, \citenamefont {Ye},
  \citenamefont {Rezaie}, \citenamefont {Yu}, \citenamefont {Liu},
  \citenamefont {He},\ and\ \citenamefont {Lui}}]{Baren2019}%
  \BibitemOpen
  \bibfield  {author} {\bibinfo {author} {\bibfnamefont {J.}~\bibnamefont {van
  Baren}}, \bibinfo {author} {\bibfnamefont {G.}~\bibnamefont {Ye}}, \bibinfo
  {author} {\bibfnamefont {J.-A.}\ \bibnamefont {Yan}}, \bibinfo {author}
  {\bibfnamefont {Z.}~\bibnamefont {Ye}}, \bibinfo {author} {\bibfnamefont
  {P.}~\bibnamefont {Rezaie}}, \bibinfo {author} {\bibfnamefont
  {P.}~\bibnamefont {Yu}}, \bibinfo {author} {\bibfnamefont {Z.}~\bibnamefont
  {Liu}}, \bibinfo {author} {\bibfnamefont {R.}~\bibnamefont {He}},\ and\
  \bibinfo {author} {\bibfnamefont {C.~H.}\ \bibnamefont {Lui}},\ }\bibfield
  {title} {\bibinfo {title} {Stacking-dependent interlayer phonons in 3{R} and
  2{H} {MoS}$_2$},\ }\href {https://doi.org/10.1088/2053-1583/ab0196}
  {\bibfield  {journal} {\bibinfo  {journal} {2D Materials}\ }\textbf {\bibinfo
  {volume} {6}},\ \bibinfo {pages} {025022} (\bibinfo {year}
  {2019})}\BibitemShut {NoStop}%
\bibitem [{Note1()}]{Note1}%
  \BibitemOpen
  \bibinfo {note} {See Supplemental Material for the methods for
  sample preparation, device fabrication, and optical spectroscopy
  measurements; the bond polarizability model for stacking-dependent shear
  modes; additional data for Device D34 and other devices; evidence for a
  prevalent chiral orientation; pulsed-voltage measurements.}\BibitemShut
  {Stop}%
\bibitem [{\citenamefont {Fan}\ \emph {et~al.}(2025)\citenamefont {Fan},
  \citenamefont {Zhang}, \citenamefont {Yang}, \citenamefont {Li},
  \citenamefont {Li}, \citenamefont {Zhang}, \citenamefont {Gao}, \citenamefont
  {Wu}, \citenamefont {Wu}, \citenamefont {Geng},\ and\ \citenamefont
  {Hu}}]{Fan2025}%
  \BibitemOpen
  \bibfield  {author} {\bibinfo {author} {\bibfnamefont {A.}~\bibnamefont
  {Fan}}, \bibinfo {author} {\bibfnamefont {Q.}~\bibnamefont {Zhang}}, \bibinfo
  {author} {\bibfnamefont {Z.}~\bibnamefont {Yang}}, \bibinfo {author}
  {\bibfnamefont {L.}~\bibnamefont {Li}}, \bibinfo {author} {\bibfnamefont
  {M.}~\bibnamefont {Li}}, \bibinfo {author} {\bibfnamefont {K.}~\bibnamefont
  {Zhang}}, \bibinfo {author} {\bibfnamefont {J.}~\bibnamefont {Gao}}, \bibinfo
  {author} {\bibfnamefont {F.}~\bibnamefont {Wu}}, \bibinfo {author}
  {\bibfnamefont {M.}~\bibnamefont {Wu}}, \bibinfo {author} {\bibfnamefont
  {D.}~\bibnamefont {Geng}},\ and\ \bibinfo {author} {\bibfnamefont
  {W.}~\bibnamefont {Hu}},\ }\bibfield  {title} {\bibinfo {title} {Tailored
  sliding ferroelectricity for ultrahigh fatigue resistance in stacked trilayer
  {MoS$_2$} crystals},\ }\href {https://doi.org/10.1126/sciadv.adx8192}
  {\bibfield  {journal} {\bibinfo  {journal} {Sci. Adv.}\ }\textbf {\bibinfo
  {volume} {11}},\ \bibinfo {pages} {eadx8192} (\bibinfo {year}
  {2025})}\BibitemShut {NoStop}%
\bibitem [{\citenamefont {Butz}\ \emph {et~al.}(2014)\citenamefont {Butz},
  \citenamefont {Dolle}, \citenamefont {Niekiel}, \citenamefont {Weber},
  \citenamefont {Waldmann}, \citenamefont {Weber}, \citenamefont {Meyer},\ and\
  \citenamefont {Spiecker}}]{Butz2014}%
  \BibitemOpen
  \bibfield  {author} {\bibinfo {author} {\bibfnamefont {B.}~\bibnamefont
  {Butz}}, \bibinfo {author} {\bibfnamefont {C.}~\bibnamefont {Dolle}},
  \bibinfo {author} {\bibfnamefont {F.}~\bibnamefont {Niekiel}}, \bibinfo
  {author} {\bibfnamefont {K.}~\bibnamefont {Weber}}, \bibinfo {author}
  {\bibfnamefont {D.}~\bibnamefont {Waldmann}}, \bibinfo {author}
  {\bibfnamefont {H.~B.}\ \bibnamefont {Weber}}, \bibinfo {author}
  {\bibfnamefont {B.}~\bibnamefont {Meyer}},\ and\ \bibinfo {author}
  {\bibfnamefont {E.}~\bibnamefont {Spiecker}},\ }\bibfield  {title} {\bibinfo
  {title} {Dislocations in bilayer graphene},\ }\href
  {https://doi.org/10.1038/nature12780} {\bibfield  {journal} {\bibinfo
  {journal} {Nature}\ }\textbf {\bibinfo {volume} {505}},\ \bibinfo {pages}
  {533} (\bibinfo {year} {2014})}\BibitemShut {NoStop}%
\bibitem [{\citenamefont {Yankowitz}\ \emph {et~al.}(2014)\citenamefont
  {Yankowitz}, \citenamefont {Wang}, \citenamefont {Birdwell}, \citenamefont
  {Chen}, \citenamefont {Watanabe}, \citenamefont {Taniguchi}, \citenamefont
  {Jacquod}, \citenamefont {San-Jose}, \citenamefont {Jarillo-Herrero},\ and\
  \citenamefont {LeRoy}}]{Yankowitz2014}%
  \BibitemOpen
  \bibfield  {author} {\bibinfo {author} {\bibfnamefont {M.}~\bibnamefont
  {Yankowitz}}, \bibinfo {author} {\bibfnamefont {J.~I.~J.}\ \bibnamefont
  {Wang}}, \bibinfo {author} {\bibfnamefont {A.~G.}\ \bibnamefont {Birdwell}},
  \bibinfo {author} {\bibfnamefont {Y.-A.}\ \bibnamefont {Chen}}, \bibinfo
  {author} {\bibfnamefont {K.}~\bibnamefont {Watanabe}}, \bibinfo {author}
  {\bibfnamefont {T.}~\bibnamefont {Taniguchi}}, \bibinfo {author}
  {\bibfnamefont {P.}~\bibnamefont {Jacquod}}, \bibinfo {author} {\bibfnamefont
  {P.}~\bibnamefont {San-Jose}}, \bibinfo {author} {\bibfnamefont
  {P.}~\bibnamefont {Jarillo-Herrero}},\ and\ \bibinfo {author} {\bibfnamefont
  {B.~J.}\ \bibnamefont {LeRoy}},\ }\bibfield  {title} {\bibinfo {title}
  {Electric field control of soliton motion and stacking in trilayer
  graphene},\ }\href {https://doi.org/10.1038/nmat3965} {\bibfield  {journal}
  {\bibinfo  {journal} {Nat. Mater.}\ }\textbf {\bibinfo {volume} {13}},\
  \bibinfo {pages} {786} (\bibinfo {year} {2014})}\BibitemShut {NoStop}%
\bibitem [{\citenamefont {Jiang}\ \emph {et~al.}(2018)\citenamefont {Jiang},
  \citenamefont {Wang}, \citenamefont {Shi}, \citenamefont {Jin}, \citenamefont
  {Utama}, \citenamefont {Zhao}, \citenamefont {Shen}, \citenamefont {Gao},
  \citenamefont {Zhang},\ and\ \citenamefont {Wang}}]{Jiang2018}%
  \BibitemOpen
  \bibfield  {author} {\bibinfo {author} {\bibfnamefont {L.}~\bibnamefont
  {Jiang}}, \bibinfo {author} {\bibfnamefont {S.}~\bibnamefont {Wang}},
  \bibinfo {author} {\bibfnamefont {Z.}~\bibnamefont {Shi}}, \bibinfo {author}
  {\bibfnamefont {C.}~\bibnamefont {Jin}}, \bibinfo {author} {\bibfnamefont
  {M.~I.~B.}\ \bibnamefont {Utama}}, \bibinfo {author} {\bibfnamefont
  {S.}~\bibnamefont {Zhao}}, \bibinfo {author} {\bibfnamefont {Y.-R.}\
  \bibnamefont {Shen}}, \bibinfo {author} {\bibfnamefont {H.-J.}\ \bibnamefont
  {Gao}}, \bibinfo {author} {\bibfnamefont {G.}~\bibnamefont {Zhang}},\ and\
  \bibinfo {author} {\bibfnamefont {F.}~\bibnamefont {Wang}},\ }\bibfield
  {title} {\bibinfo {title} {Manipulation of domain-wall solitons in bi- and
  trilayer graphene},\ }\href {https://doi.org/10.1038/s41565-017-0042-6}
  {\bibfield  {journal} {\bibinfo  {journal} {Nat. Nanotechnol.}\ }\textbf
  {\bibinfo {volume} {13}},\ \bibinfo {pages} {204} (\bibinfo {year}
  {2018})}\BibitemShut {NoStop}%
\bibitem [{\citenamefont {Zhang}\ \emph {et~al.}(2022)\citenamefont {Zhang},
  \citenamefont {Xu}, \citenamefont {Hou}, \citenamefont {Song}, \citenamefont
  {Ma}, \citenamefont {Gao}, \citenamefont {Zhu}, \citenamefont {Ma},
  \citenamefont {Liu}, \citenamefont {Feng},\ and\ \citenamefont
  {Li}}]{Zhang2022}%
  \BibitemOpen
  \bibfield  {author} {\bibinfo {author} {\bibfnamefont {S.}~\bibnamefont
  {Zhang}}, \bibinfo {author} {\bibfnamefont {Q.}~\bibnamefont {Xu}}, \bibinfo
  {author} {\bibfnamefont {Y.}~\bibnamefont {Hou}}, \bibinfo {author}
  {\bibfnamefont {A.}~\bibnamefont {Song}}, \bibinfo {author} {\bibfnamefont
  {Y.}~\bibnamefont {Ma}}, \bibinfo {author} {\bibfnamefont {L.}~\bibnamefont
  {Gao}}, \bibinfo {author} {\bibfnamefont {M.}~\bibnamefont {Zhu}}, \bibinfo
  {author} {\bibfnamefont {T.}~\bibnamefont {Ma}}, \bibinfo {author}
  {\bibfnamefont {L.}~\bibnamefont {Liu}}, \bibinfo {author} {\bibfnamefont
  {X.-Q.}\ \bibnamefont {Feng}},\ and\ \bibinfo {author} {\bibfnamefont
  {Q.}~\bibnamefont {Li}},\ }\bibfield  {title} {\bibinfo {title} {Domino-like
  stacking order switching in twisted monolayer–multilayer graphene},\ }\href
  {https://doi.org/10.1038/s41563-022-01232-2} {\bibfield  {journal} {\bibinfo
  {journal} {Nat. Mater.}\ }\textbf {\bibinfo {volume} {21}},\ \bibinfo {pages}
  {621} (\bibinfo {year} {2022})}\BibitemShut {NoStop}%
\bibitem [{\citenamefont {Malard}\ \emph {et~al.}(2013)\citenamefont {Malard},
  \citenamefont {Alencar}, \citenamefont {Barboza}, \citenamefont {Mak},\ and\
  \citenamefont {de~Paula}}]{Malard2013}%
  \BibitemOpen
  \bibfield  {author} {\bibinfo {author} {\bibfnamefont {L.~M.}\ \bibnamefont
  {Malard}}, \bibinfo {author} {\bibfnamefont {T.~V.}\ \bibnamefont {Alencar}},
  \bibinfo {author} {\bibfnamefont {A.~P.~M.}\ \bibnamefont {Barboza}},
  \bibinfo {author} {\bibfnamefont {K.~F.}\ \bibnamefont {Mak}},\ and\ \bibinfo
  {author} {\bibfnamefont {A.~M.}\ \bibnamefont {de~Paula}},\ }\bibfield
  {title} {\bibinfo {title} {Observation of intense second harmonic generation
  from {MoS}$_2$ atomic crystals},\ }\href
  {https://doi.org/10.1103/PhysRevB.87.201401} {\bibfield  {journal} {\bibinfo
  {journal} {Phys. Rev. B}\ }\textbf {\bibinfo {volume} {87}},\ \bibinfo
  {pages} {201401} (\bibinfo {year} {2013})}\BibitemShut {NoStop}%
\bibitem [{\citenamefont {Dresselhaus}\ and\ \citenamefont
  {Avouris}(2001)}]{Dresselhaus2001}%
  \BibitemOpen
  \bibfield  {author} {\bibinfo {author} {\bibfnamefont {M.~S.}\ \bibnamefont
  {Dresselhaus}}\ and\ \bibinfo {author} {\bibfnamefont {P.}~\bibnamefont
  {Avouris}},\ }\bibinfo {title} {Introduction to carbon materials research},\
  in\ \href {https://doi.org/10.1007/3-540-39947-X_1} {\emph {\bibinfo
  {booktitle} {Carbon Nanotubes: Synthesis, Structure, Properties, and
  Applications}}},\ \bibinfo {editor} {edited by\ \bibinfo {editor}
  {\bibfnamefont {M.~S.}\ \bibnamefont {Dresselhaus}}, \bibinfo {editor}
  {\bibfnamefont {G.}~\bibnamefont {Dresselhaus}},\ and\ \bibinfo {editor}
  {\bibfnamefont {P.}~\bibnamefont {Avouris}}}\ (\bibinfo  {publisher}
  {Springer Berlin Heidelberg},\ \bibinfo {address} {Berlin, Heidelberg},\
  \bibinfo {year} {2001})\ pp.\ \bibinfo {pages} {1--9}\BibitemShut {NoStop}%
\bibitem [{\citenamefont {Guo}\ \emph {et~al.}(2016)\citenamefont {Guo},
  \citenamefont {Liu}, \citenamefont {Yin}, \citenamefont {Wei}, \citenamefont
  {Lin}, \citenamefont {Hoffman}, \citenamefont {Zhao}, \citenamefont {Edgar},
  \citenamefont {Chen}, \citenamefont {Lau}, \citenamefont {Dai}, \citenamefont
  {Yao}, \citenamefont {Wong},\ and\ \citenamefont {Chai}}]{Guo2016}%
  \BibitemOpen
  \bibfield  {author} {\bibinfo {author} {\bibfnamefont {Y.}~\bibnamefont
  {Guo}}, \bibinfo {author} {\bibfnamefont {C.}~\bibnamefont {Liu}}, \bibinfo
  {author} {\bibfnamefont {Q.}~\bibnamefont {Yin}}, \bibinfo {author}
  {\bibfnamefont {C.}~\bibnamefont {Wei}}, \bibinfo {author} {\bibfnamefont
  {S.}~\bibnamefont {Lin}}, \bibinfo {author} {\bibfnamefont {T.~B.}\
  \bibnamefont {Hoffman}}, \bibinfo {author} {\bibfnamefont {Y.}~\bibnamefont
  {Zhao}}, \bibinfo {author} {\bibfnamefont {J.~H.}\ \bibnamefont {Edgar}},
  \bibinfo {author} {\bibfnamefont {Q.}~\bibnamefont {Chen}}, \bibinfo {author}
  {\bibfnamefont {S.~P.}\ \bibnamefont {Lau}}, \bibinfo {author} {\bibfnamefont
  {J.}~\bibnamefont {Dai}}, \bibinfo {author} {\bibfnamefont {H.}~\bibnamefont
  {Yao}}, \bibinfo {author} {\bibfnamefont {H.-S.~P.}\ \bibnamefont {Wong}},\
  and\ \bibinfo {author} {\bibfnamefont {Y.}~\bibnamefont {Chai}},\ }\bibfield
  {title} {\bibinfo {title} {Distinctive in-plane cleavage behaviors of
  two-dimensional layered materials},\ }\href
  {https://doi.org/10.1021/acsnano.6b05063} {\bibfield  {journal} {\bibinfo
  {journal} {ACS Nano}\ }\textbf {\bibinfo {volume} {10}},\ \bibinfo {pages}
  {8980} (\bibinfo {year} {2016})}\BibitemShut {NoStop}%
\bibitem [{\citenamefont {Yeo}\ \emph {et~al.}(2025)\citenamefont {Yeo},
  \citenamefont {Sharaby}, \citenamefont {Roy}, \citenamefont {Raab},
  \citenamefont {Watanabe}, \citenamefont {Taniguchi},\ and\ \citenamefont
  {Ben~Shalom}}]{Yeo2025}%
  \BibitemOpen
  \bibfield  {author} {\bibinfo {author} {\bibfnamefont {Y.}~\bibnamefont
  {Yeo}}, \bibinfo {author} {\bibfnamefont {Y.}~\bibnamefont {Sharaby}},
  \bibinfo {author} {\bibfnamefont {N.}~\bibnamefont {Roy}}, \bibinfo {author}
  {\bibfnamefont {N.}~\bibnamefont {Raab}}, \bibinfo {author} {\bibfnamefont
  {K.}~\bibnamefont {Watanabe}}, \bibinfo {author} {\bibfnamefont
  {T.}~\bibnamefont {Taniguchi}},\ and\ \bibinfo {author} {\bibfnamefont
  {M.}~\bibnamefont {Ben~Shalom}},\ }\bibfield  {title} {\bibinfo {title}
  {Polytype switching by super-lubricant van der {W}aals cavity arrays},\
  }\href {https://doi.org/10.1038/s41586-024-08380-2} {\bibfield  {journal}
  {\bibinfo  {journal} {Nature}\ }\textbf {\bibinfo {volume} {638}},\ \bibinfo
  {pages} {389} (\bibinfo {year} {2025})}\BibitemShut {NoStop}%
\bibitem [{\citenamefont {Liu}\ \emph {et~al.}(2026)\citenamefont {Liu},
  \citenamefont {Watanabe}, \citenamefont {Taniguchi},\ and\ \citenamefont
  {Xi}}]{DAS}%
  \BibitemOpen
  \bibfield  {author} {\bibinfo {author} {\bibfnamefont {Y.}~\bibnamefont
  {Liu}}, \bibinfo {author} {\bibfnamefont {K.}~\bibnamefont {Watanabe}},
  \bibinfo {author} {\bibfnamefont {T.}~\bibnamefont {Taniguchi}},\ and\
  \bibinfo {author} {\bibfnamefont {X.}~\bibnamefont {Xi}},\ }\href
  {https://doi.org/10.6084/m9.figshare.32114560} {\bibinfo {title} {Data for
  ``{Shear-mode Raman Imaging of Ferroelectric Switching in Multilayer
  3$R$-MoS$_2$}''}} (\bibinfo {year} {2026})\BibitemShut {NoStop}%
\end{thebibliography}
\end{document}